\documentclass{aa}
\usepackage[utf8]{inputenc}
\usepackage{ngerman}
\usepackage{verbatim}
\usepackage{amsmath}
\usepackage{xcolor}
\usepackage{mathcomp}
\usepackage{float}
\usepackage{array}
\usepackage{graphicx}
\usepackage{geometry}
\usepackage{url}
\usepackage{hyperref} 
\usepackage{attachfile2}
\usepackage{tablefootnote}
\usepackage[separate-uncertainty]{siunitx}
\usepackage[varg]{txfonts}
\usepackage{ragged2e}
\bibpunct{(}{)}{;}{a}{}{,}
\renewcommand{\d}{\ensuremath{\operatorname{d}\!}}
\newcommand{\derive}[2]{\ensuremath{\dfrac{\d #1}{\d #2}}}
\newcommand{\partdiff}[2]{\ensuremath{\dfrac{\partial \! #1}{\partial \! #2}}}

\date{Received date /Accepted date }
\title{Thermal evolution of Uranus and Neptune II - Deep thermal boundary layer}
\author{Ludwig Scheibe\inst{1} \and Nadine Nettelmann\inst{2} \and Ronald Redmer\inst{1}}
\institute{Institut für Physik, Universität Rostock, A.-Einstein-Str. 23, D-18059 Rostock, Germany \and Institut für Planetenforschung, Deutsches Zentrum für Luft- und Raumfahrt,  Rutherfordstraße 2, D-12489 Berlin, Germany}
\abstract{
Thermal evolution models suggest that the luminosities of both Uranus and Neptune are inconsistent with the classical assumption of an adiabatic interior. Such models commonly predict Uranus to be brighter and, recently, Neptune to be fainter than observed. In this work, we investigate the influence of a thermally conductive boundary layer on the evolution of Uranus- and Neptune-like planets. This thermal boundary layer (TBL) is assumed to be located deep in the planet and be caused by a steep compositional gradient between a H-He-dominated outer envelope and an ice-rich inner envelope. We investigate the effect of TBL thickness, thermal conductivity, and the time of TBL formation on the planet's cooling behaviour.\\
The calculations were performed with our recently developed tool based on the Henyey method for stellar evolution. We make use of state-of-the-art equations of state for hydrogen, helium, and water, as well as of thermal conductivity data for water calculated via \textit{ab initio} methods. \\
We find that even a thin conductive layer of a few kilometres has a significant influence on the planetary cooling.
In our models, Uranus' measured luminosity can only be reproduced if the planet has been near equilibrium with the solar incident flux for an extended period of time. For Neptune, we find a range of solutions with a near constant effective temperature at layer thicknesses of 15 km or larger, similar to Uranus. In addition, we find solutions for thin TBLs of a few km and strongly enhanced thermal conductivity. A $\sim$\SI{1}{Gyr} later  onset of the TBL reduces the present $\Delta T$ by an order of magnitude to only several \SI{100}{K}.\\ 
Our models suggest that a TBL can  significantly influence the present planetary luminosity in both directions, making it appear either brighter or fainter than the adiabatic case.
}
\usepackage[normalem]{ulem}

\begin{document}

\maketitle
\section{Introduction}
\label{sec:intro}
Reconciling the perceived similarities between Uranus and Neptune, with their highly different intrinsic heat fluxes, is a long-standing challenge. While the two planets display a number of similar key quantities, such as mass, radius, surface temperature, and atmospheric composition \citep{Guillot15}, as well as non-dipolar magnetic fields \citep{Connerney1991}, Neptune's intrinsic heat flux of $F_\text{N} = \SI{0.433 \pm 0.046}{\watt \metre ^{-2}}$ is about an order of magnitude higher than that of Uranus, with $F_\text{U}=0.042_{-0.042}^{+0.047}\: \si{\watt \metre ^{-2}}$  \citep{Guillot15}. For Uranus, measurement uncertainties in Albedo and emitted flux  would allow for the net heat loss to be zero. Classical evolution calculations assuming an interior that is fully convective and thus adiabatic yield cooling times for Uranus that are significantly larger than the age of the Solar System \citep{Podolak1990, Fortney11, Nettelmann13, Linder19, Scheibe19}. Neptune's cooling time has previously been predicted to be in accordance with the age of the Solar System \citep{Fortney11, Nettelmann13, Linder19}; however, using more up-to-date equations of state, we find it to be too short (\citealp[][]{Scheibe19}, hereafter Paper I). A review of this problem can be found in \citet{Helled20}. \\
Several proposals have recently been put forward to explain Uranus' low heat flow.  
\citet{Kurosaki17} argue that condensation of the volatiles H$_2$O, CH$_4$, and NH$_3$ at around the 1-bar to 100-bar levels leads to latent heat release into the atmosphere. As a result, the radiative atmosphere above becomes warmer and can lose heat more efficiently. This would speed up the planetary cooling at young ages and let the planet appear brighter when young and fainter when old. However, such a process would require a larger amount of condensibles in the outer envelope of Uranus than consistent with interior model predictions. \citet{Vazan20} investigated a primordial composition gradient in the fluid envelope inhibiting convective energy transport. They find such a gradient to be stable and were able to reproduce Uranus' observed luminosity with these models. 
Another possible explanation is that the water in the deep interior transitions into the superionic phase, shutting a growing proportion of the planet off from efficient cooling because of the high viscosity in a superionic interior \citep{Stixrude20}. Neptune, on the other hand, has not received nearly as much attention. \\
In this work, we considered the thermal evolution of both Uranus and Neptune under the assumption of a single thermal boundary layer (TBL) in the envelope as a result of a steep compositional gradient between the ice-rock-rich deep interior and the hydrogen- or helium-dominated outer envelope. This boundary would trap part of the primordial heat inside, allowing the outer envelope to cool quicker than in the adiabatic case, an idea put forward for Uranus and Neptune by \citet{Hubbard95}.\\
Similarly to \citet{Nettelmann16}, who have previously investigated this possibility for Uranus, we assumed the simplest type of internal structure that is consistent with the gravity data, namely a  three-layer structure of a rocky core, ice-rich inner envelope and H/He-rich outer envelope. \citet{Nettelmann16} used ad hoc functional forms for the evolution of the temperature gradient across the TBL that would force it to strengthen with time, an assumption sufficient to explain the low luminosity of Uranus but not the brightness of Neptune. \citet{Helled20} showed that including the possibility of the TBL decaying in this setup significantly broadens the range of present luminosities and might even produce higher present luminosity than the adiabatic case, which might help explain the brightness of Neptune. Periods of convective instability erroding the TBL might indeed occur when the thermal buoyancy induced by the TBL overcomes the stabilising effect of the compositional gradient \citep{Vazan20}. It is also possible that the inhomogenous structure of the ice giants inferred for present Uranus and Neptune \citep{Helled20b} is the result of more recent phase separation and sedimentation processes. \citet{Bailey21} extrapolated experimental data on H$_2$-H$_2$O demixing to the interior conditions of adiabatic Uranus and Neptune to argue that this process might play a significant role in shaping their composition distribution and  evolution of their heat flow. In this picture, H$_2$-H$_2$O demixing would actively occur in present Neptune and slow down its cooling rate, while in Uranus the demixing should be closer to completion to explain its faintness as well as lower outer envelope metallicity inferred from modelling.  However, a homogeneous interior prior to subsequent water sedimentation might lead to a higher-than-observed D/H ratio unless the interior is rock rich rather than ice rich \citep{Feuchtgruber13}, suggesting that primordial composition gradients were never fully eroded. On top of that, condensation of abundant water in the deep atmosphere can also significantly affect the heat transport mechanism and thus the surface luminosity \citep{Leconte17,Friedson17}. \\ 
Our models do not aim to represent the real and certainly more complex structure  of Uranus and Neptune. Our goal is rather to systematically study what influence a TBL has on the luminosity evolution of a planet with the mass, size, and orbital parameters of Uranus and Neptune, whatever processes may act upon the planet in addition to that.\\
Because of the narrow window for semi-convective conditions in Uranus and Neptune \citep{French19b} and the transient nature of convective erosion \citep{Vazan20}, we assume a static, conductive TBL. Uncertainties on the thermal state are included by varying the base thermal conductivity, its onset after formation, and its thickness. We used thermal conductivity values for water calculated via DFT-MD \citep{French17,French19}.\\
In Sect. \ref{sec:methods}, we describe our numerical method and the model assumptions; in Sect. \ref{sec:results}, we present our results and discuss them; and Sect. \ref{sec:conclusions} summarises our conclusions.%
\section{Methodology}
\label{sec:methods}
\subsection{Theoretical foundations}
\label{sec:met_theory}
To compute the thermal evolution models, we used the OTTER tool (Paper I). It accounts for conservation of mass, hydrostatic equilibrium, energy transport, and conservation of energy for the main variables of radius $r(m)$, pressure $P(m)$, temperature $T(m)$, and luminosity $l(m)$ \mbox{\citep{KippWei}}, taking into account solid-body rotation as a spherically symmetric zeroth-order approximation. In particular, the internal temperature profile is given by
  \begin{align}
          \dfrac{\partial  T}{\partial m} = \left(  -\frac{G\,m}{4\pi\, r^4} + \frac{\omega^2}{6\pi\, r } \right) \frac{T}{P}\nabla_{T}, \label{eq:Tm}
  \end{align}
where $\omega$ is the planet's rotation rate and $\nabla_T=\dfrac{\partial  \ln T}{\partial \ln P}$ denotes the temperature gradient that is evaluated at each point separately. In this work, the adiabatic and conductive temperature gradients are used, which can be formulated as follows \citep{KippWei}:
\begin{subequations}
\label{eq:nabla}
  \begin{align}
    \nabla_{\text{ad}}&=\left( \dfrac{\partial  \ln T}{\partial \ln P} \right)_S=\frac{P\: \delta}{T\:\rho\: c_p}, \label{eq:nabla_ad}\\
    \nabla_{ \text{cond}} &= \frac{1}{4 \pi} \frac{l\: P}{G\: m\: T\: \rho\: \lambda}, \label{eq:nabla_con}
  \end{align}
\end{subequations}
where $\delta=-\left( \dfrac{\partial  \ln\rho}{\partial \ln T} \right)_P$ is the thermal expansion coefficient, $c_p=\left( \dfrac{\partial  u}{\partial T} \right)_P - \frac{P}{\rho^2} \left( \dfrac{\partial  \rho}{\partial T} \right)_P$ the isobaric heat capacity, and $\lambda$ the thermal conductivity. 
Because our models feature an approximation of rotation in zeroth order, Eq. \eqref{eq:nabla_con} is a simplification of the actual gradient. However, since the additional term due to ration is smaller than $\SI{1.5}{\%}$ for the TBL regions in all our models and its influence on the cooling times is insignificant, we neglect it in this description.\\
The luminosity profile is obtained by integrating the following equation over mass:
 \begin{equation}
     \dfrac{\partial  l}{\partial m} = - c_p \dfrac{\partial  T}{\partial t} + \frac{\delta}{\rho} \dfrac{\partial  P}{\partial t}.\label{eq:lm}
 \end{equation}
The two outer boundary conditions at $m=M_\text{P}$ are 
\begin{subequations}
  \begin{align}
    P (M_\text{P})&=\SI{1}{\bar}, \label{eq:B0}\\
    l (M_\text{P})&=4 \pi R^2 \sigma_\text{B} T_\text{eff}^4 - L_\text{sol}, \label{eq:B1}
  \end{align}
\end{subequations}
where $L_\text{sol}$ is the solar radiation absorbed and re-emitted by the planet and $T_\text{eff}$ denotes the planet's effective temperature linked to the one-bar temperature and gravity $g=GM/R^2$ via 
\begin{equation}
  T_\text{1bar}=K g^{-1/6} T_\text{eff}^{1.244}, \label{eq:1bar_eff}
\end{equation}
which is based on model atmospheres for Jupiter developed by \citet{Grab75} and interpolated by \citet{Hubb77}. The parameter $K$ is empirically chosen to yield the observed one-bar
temperatures for Uranus and Neptune for their present-day radii and effective temperatures as $K_\text{U}=1.481$ for Uranus and $K_\text{N}=1.451$ for Neptune. Both Uranus and Neptune models use a Bond albedo value of $A = 0.3$, in accordance with the measured values \citep{Pearl90, Pearl91}. For a given time, the structure equations are solved iteratively from an estimated solution using Newton's method, and then a new solution for the next time step is estimated based on the current and the previous time step. This gives a succession of profiles that are linked via the time derivatives in Eq. \eqref{eq:lm}, and from these profiles the evolution curves of $R(t)$, $T_\text{eff}(t)$, etc. can be constructed. Further details concerning the implementation can be found in Paper I.
\subsection{Model assumptions}
\label{sec:met_model}
\begin{figure}
    \centering
    \includegraphics[width=0.38\textwidth]{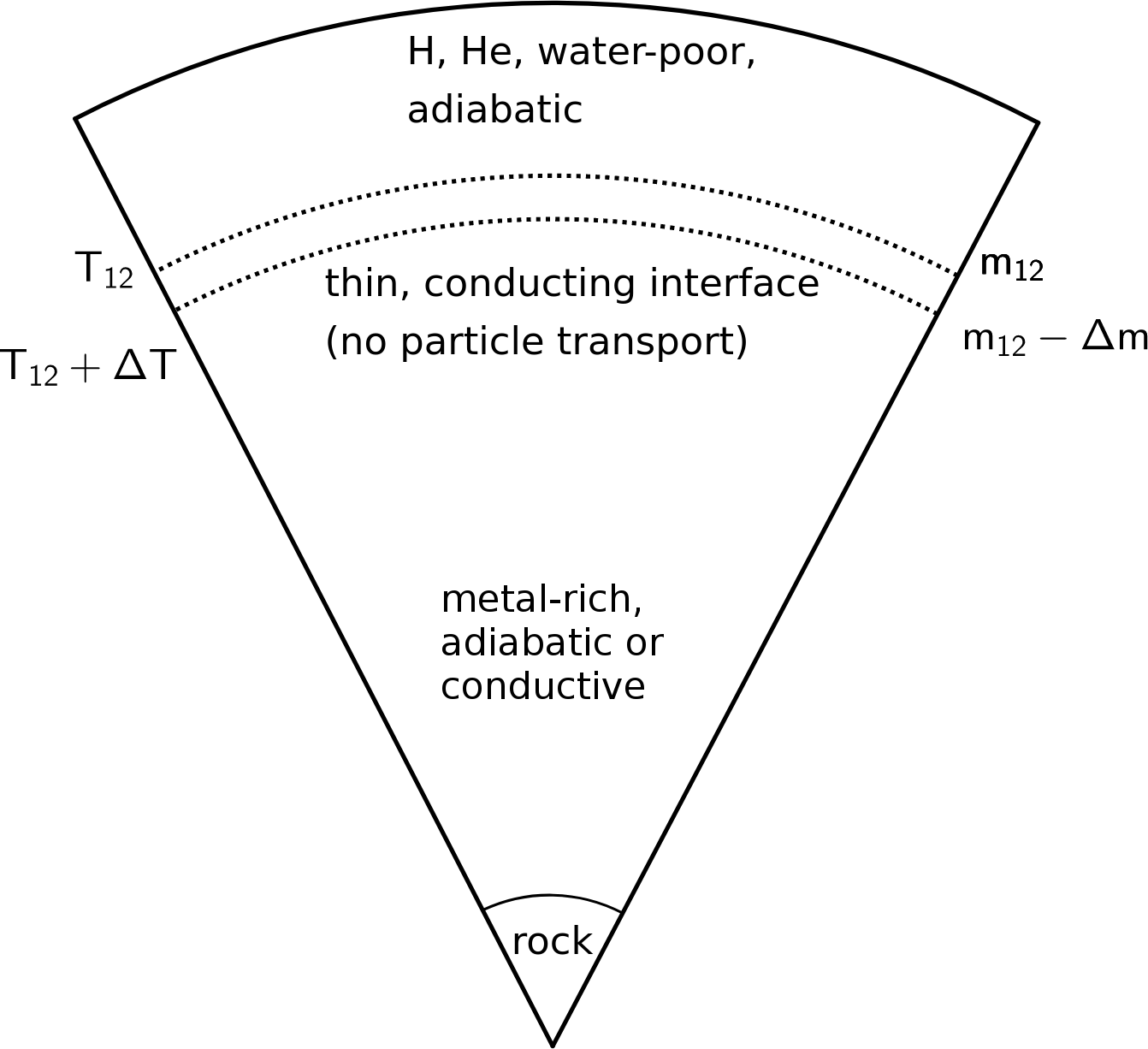}
    \caption{Sketch of the assumed three-layer structure: Isothermal rock core surrounded by heavy-element-rich inner envelope, in turn surrounded by a (H,He)-rich outer envelope. Between the inner and outer envelopes, a thermally conducting TBL of thickness $\Delta m$ is assumed, which leads to a temperature difference of $\Delta T$ across the TBL. }
    \label{fig:Modelltorte}
\end{figure}
We assumed a three-layer-structure as presented in, for example, \citet{Podolak95} or \citet{Nettelmann13} as a baseline. A small isothermal rock core is surrounded by a uniform inner envelope of heavy elements, which are represented by an equation of state (EOS) for H$_2$O and, in some cases, basalt. This ice layer is surrounded by an outer envelope of H, He, and H$_2$O. The outer envelope is considered to be vigorously convecting and thus adiabatic, while the inner envelope assumes the lower of the conducting and adiabatic temperature gradients, following the Schwarzschild criterion for convection. However, contrary to the classical setup, these two layers are separated by an interface {-- called the TBL --} in which the composition is set to form a linear gradient of $X_i$ with regard to mass $m$, where $X_i$ are the different species' mass fractions, from the values in the outer to the inner envelope and in which the temperature gradient follows $\nabla_\text{cond}$. The transition from the outer envelope to the TBL occurs at a defined mass $m_{12}$ and the TBL has a thickness of $\Delta m$ in mass that is kept constant throughout the evolution. 
All our evolution tracks begin at $t=0$ with a hot adiabatic profile of $T_\text{1bar}=\SI{700}{K}$. Because of the different compositions, this means that Uranus and Neptune models start at different radii and thus different effective temperatures. The starting temperature affects the initial energy content of the planets and thus the long-term evolution.  
However, as the exploration of different initial states is beyond the scope of the present work, we take this simplified approach of a universal starting 1-bar temperature.\\ 
Our choice of \SI{700}{K} yields initial luminosities after the first time step of {$L_\text{P} \sim \SI{5e18}{W}$} or $L_\text{P} \sim \num{1.3e-8}~L_\odot$. This is near the lower limit of initial luminosities of H/He planets predicted by cold start core accretion formation. By extrapolating the lower boundary of the corridor of luminosities at 3~Myr in Figure~2 of \citep{Mordasini17}, which has a slope $\Delta \log(L_\text{P}/L_\odot)$ over $\Delta \log (M_\text{P}/M_\text{Jup})$ of $\sim 20$, down to $M_p=0.05 M_\text{Jup}$, we obtain $L_\text{P}/L_\odot \sim \num{5e-8}$ as a prediction of the luminosity of $\sim 3$ -Myr-young Uranus and Neptune, which is close to our initial condition. \\
\begin{figure}
    \centering
    \includegraphics[width=0.49\textwidth]{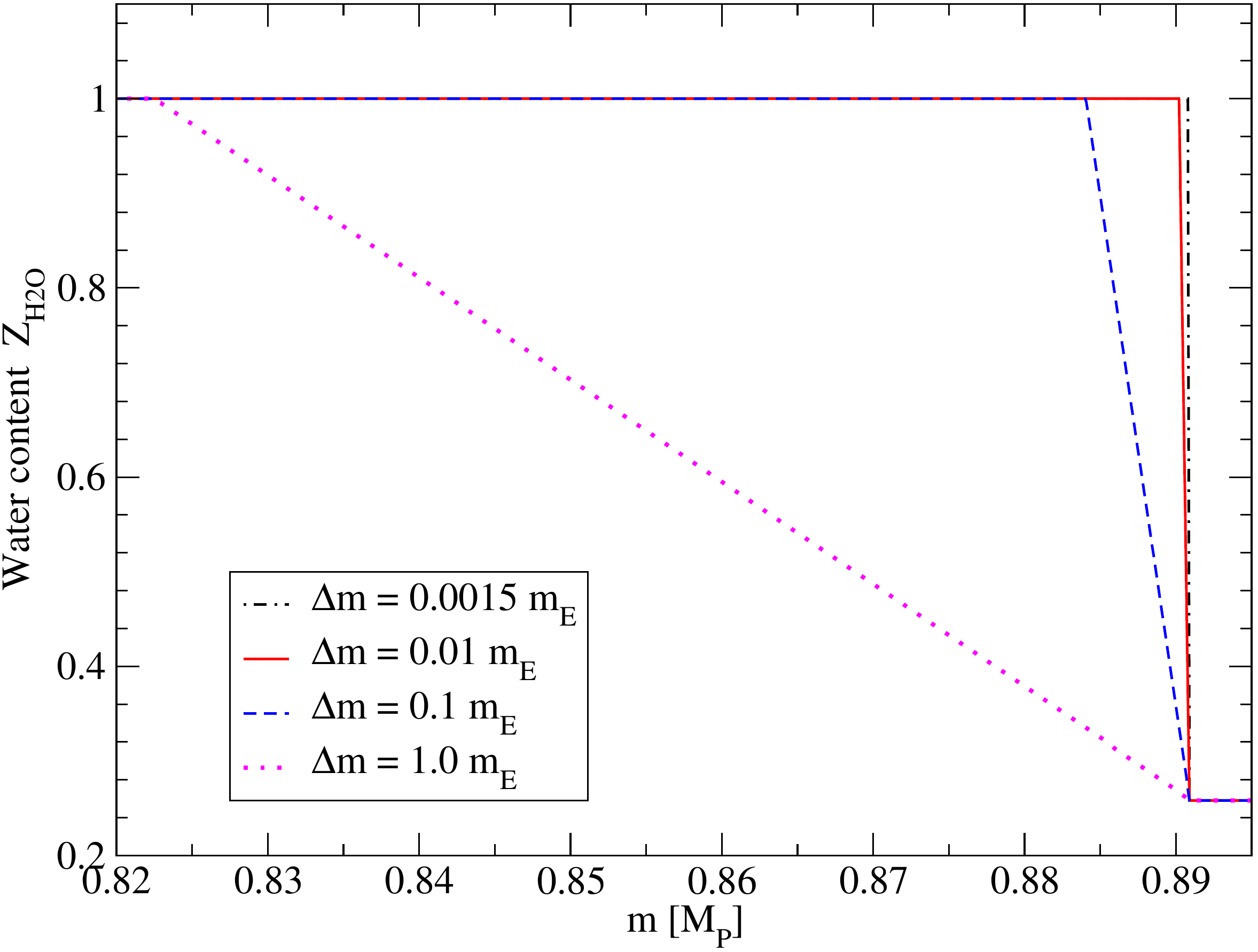}
    \caption{Water content values $Z_\text{H2O}$ along Uranus profiles with a conducting TBL of varying thickness $\Delta m$. The model makes the $Z$-gradient for high $\Delta m$ shallower than for low $\Delta m$. $Z$ outside the TBL region is constant.}
    \label{fig:Zm}
\end{figure}
Figure~\ref{fig:Modelltorte} illustrates the assumed structure, while Table\ \ref{tab:key_numbers} shows the key model parameters used for Uranus and Neptune in this work. 
Because the composition is set to a linear gradient in the TBL, higher $\Delta m$ values lead to a shallower $Z$-gradient (see Fig. \ref{fig:Zm}). The total heavy element content $Z_\text{total}$ slightly decreases with $\Delta m$. 
Table\ \ref{tab:DeltaR} shows typical radial TBL thicknesses for some of the models, as well as their total heavy-element content.\\
The models presented here are simplifications and take only some of the observed properties of the real ice giants into consideration. While we account for mass, present radius, and present luminosity, we do not aim to reproduce the gravitational moments $J_2, J_4$. However, the three-layer structure assumption together with the chosen metallicities, core mass, and transition mass $m_{12}$ closely resembles models that do fit $J_2$, $J_4$ \citep{Nettelmann13}. 
Furthermore, we do not address the inner dynamics of the planet such as possible mixing and demixing of the materials and differentiation occurring as a result of this or convective mixing. However, while these models are not accurate depictions of the real ice giants, we nevertheless call them 'Uranus' and 'Neptune', because they serve as representations. \\
In most models, the heavy elements in the inner envelope are represented entirely by water. However, in some models, in order to reproduce the planets' observed radii, additional rocky material is mixed into the inner envelope (see Sect. \ref{sec:res_Teff} for details). We assume a homogeneous mixture of water and rocks in the envelope as supported by the 1:1 water:rocks phase diagram of \citet{Vazan21Arx}. For the sake of simplicity, we considered an innermost layer of pure rock in both of these
scenarios, as shown in Fig. \ref{fig:Modelltorte}.  Uranus and Neptune models both with and without a compact rock core can satisfy the observed gravitational moments \citep{Nettelmann13}. A compact core could be a remnant of planet formation, although due to erosion it may adopt a more fuzzy state today \citep{Helled20}.\\
Hydrogen and helium are described by the Rostock EOS version 3 (REOS.3) by \citet{Becker14}, while we used a modified version of the EOS by \citet{Mazevet19} with  an ideal gas description at temperatures $T <  \SI{800}{K }$ for water. The rocky material in the core and in some models' inner envelopes is represented by the SESAME 7530 EOS for basalt \citep{Lyon92}. The different materials' EOS data are combined via the following linear mixing rule: $\rho^{-1}=\sum_i X_i/\rho_i$ and $u=\sum_iX_i u_i$.
\begin{table}
  \caption{
  Parameters used in Uranus and Neptune models here. Top half: Observational parameters taken from \citet{Guillot15} - planet mass $M_\text{p}$, mean radius $R_\text{p}$, present-day effective temperature $T_\text{eff}$ - as well as present-day equilibrium temperature $T_\text{eq}$. Bottom half: Model parameters used in this work - outer envelope metallicity $Z_1$, mass value $m_{12}$ that marks the outside edge of the TBL, and core mass $m_\text{core}$ expressed in units of Earth masses $M_\text{E}$ using a value of $M_\text{E}=\SI{5.9722e24}{kg}$. The inner envelope metallicity is $Z_2=1$ unless otherwise noted. 
  } \label{tab:key_numbers}
  \centering
  \begin{tabular}{l c c}
        \hline \hline
        & Uranus & Neptune\\
        \hline
        $M_\text{p}~/~10^{26}~\si{kg}$\footnotemark[1] & $\num{0.86832}(1\pm0.013\%)$ & $\num{1.02435}(1\pm0.013\%)$ \\ 
        $R_\text{p}~/~10^{7}~\si{m}$ & $\num{2.5364}(1\pm0.04\%)$ & $\num{2.4625}(1\pm0.08\%)$ \\
        $T_\text{eff}~/~\text{K}$ & $59.1\pm 0.3$ & $59.3\pm 0.8$ \\
        $T_\text{eq}~/~\text{K}$ & $58.1\pm 1$ & $46.4 \pm 1$ \\
        \hline
        $Z_1$ & 0.26 & 0.42 \\
        $m_{12}~/~M_\text{E}$ & 13 & 15 \\
        $m_\text{core}~/~M_\text{E}$ & 0.15 & 1 \\
        \hline \hline
  \end{tabular}
  \begin{justify}
  {\footnotesize \footnotemark[1]The given uncertainty in $M_\text{P}$ is the relative uncertainty of the gravitational constant $G$ given by \citet{Cohen87}, which was used to calculate the $M_\text{P}$ values presented here.}
  \end{justify}
\end{table}
\begin{table}

  \caption{ Typical TBL thicknesses $\Delta m$ and corresponding thicknesses in radius $\Delta r$ as well as $Z_{\rm total}$ for selected Uranus and Neptune models, assuming that the inner envelope is pure H$_2$O and that $\lambda=\lambda_\text{H2O}$. $\Delta r$ is indicated for $t=\SI{4.6}{Gyr}$; the value is between 20 and 45\% larger at the beginning of the evolution due to contraction.} \label{tab:DeltaR}
  \centering
  \begin{tabular}{l c c c c}
        \hline \hline
         & \multicolumn{2}{c}{Uranus} & \multicolumn{2}{c}{Neptune}\\
        $\Delta m$ [$M_\text{E}$] & $\Delta r$ [km]& $Z_\text{total}$ & $\Delta r$ [km] & $Z_\text{total}$\\
        \hline
        0.0015 & 3.0 & $\SI{0.9190}{ }$ & 2.2 & $\SI{0.9329}{ }$ \\
        0.005 & 10.6 & $\SI{0.9189}{ }$ & 7.4 & $\SI{0.9328}{ }$\\
        0.01 & 21.5 & $\SI{0.9188}{ }$ & 15.0 & $\SI{0.9327}{ }$\\
        0.03 & 65.2 & $\SI{0.9183}{ }$ & 45.6 & $\SI{0.9324}{ }$ \\
        1.0 & 2320 & $\SI{0.8935}{ }$ & --- & ---\\
        \hline \hline
  \end{tabular}
\end{table}
\subsection{Thermal conductivity}
\label{sec:met_conduct}
\begin{figure}
  \includegraphics[width=0.49\textwidth]{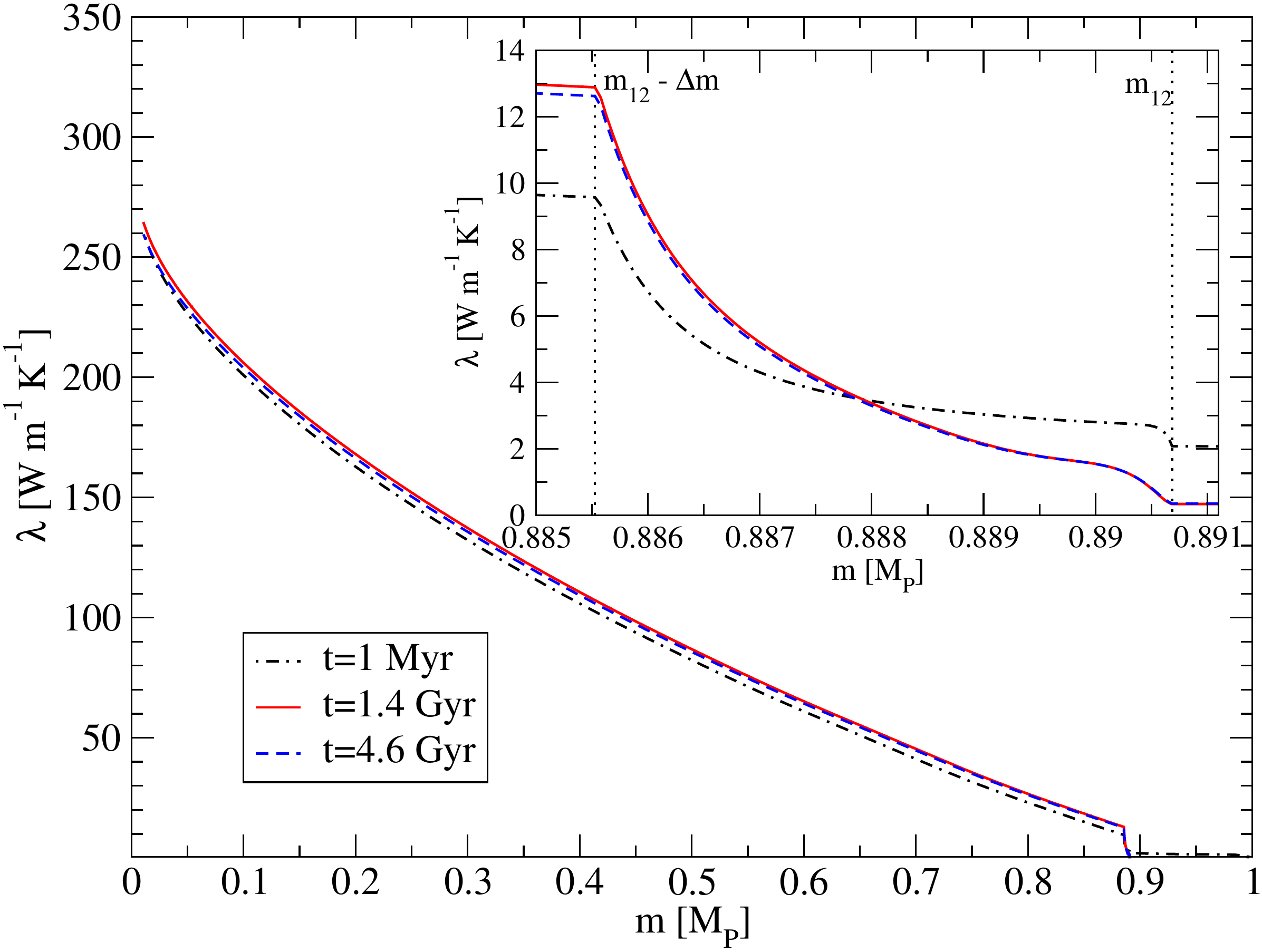}
  \caption{Thermal conductivity values along Uranus' profile with a conducting TBL of thickness of $\Delta m = 0.075~M_\text{E}$ (which corresponds to $\Delta r \approx \SI{115}{km}$ at $\SI{4.6}{Gyr}$), with an inner metallicity of $Z_{2,\text{rocks}}=0.18$ and $Z_{2,\text{H2O}}=0.82$, for three different points in the planet's lifetime at $t=\SI{1}{Myr}$, $\SI{1.4}{Gyr,}$ and $\SI{4.6}{Gyr}$. The inset shows a close-up view of the TBL region.}
  \label{fig:conduct}
\end{figure}
For the thermal conductivity $\lambda,$ we used values for pure water as presented by \cite{French17} (electronic contribution) and \cite{French19} (ionic contribution). They are based on \textit{\emph{ab initio}} DFT-MD calculations and, for the ionic contribution, have demonstrated good agreement with experimental results at and above room temperature \citep{French19} where data are available. Figure\ \ref{fig:conduct} shows the thermal conductivity along the TBL and the inner envelope of a Uranus model during the evolution at $t=\SI{1}{Myr},$ $\SI{1.4}{Gyr}$, and $\SI{4.6}{Gyr}$. This model was chosen because it reproduces both the observed $R_\text{P}$ and $T_\text{eff}$ at $t=\SI{4.6}{Gyr}$. $\lambda$ increases with both density and temperature (cf.\ Figure\ 4 in \cite{French19}), which is why $\lambda$ increases significantly towards the centre.\\
Using the values for pure water is certainly a simplification because the ices in Uranus and Neptune are more likely to be a mixture of H$_2$O, CH$_4$, NH$_3$, and refractory elements such as MgO and silicates, and H-He in the outer part of the TBL.\\ 
This approach is taken because water is assumed as the main constituent in our models and the thermal conductivity of real mixtures under the conditions relevant here are poorly understood. We used the H$_2$O values as an approximate lower bound to the real conductivity. The $\lambda$ values for hydrogen are higher than those of water under the majority the conditions of our TBL models \citep{Holst11}.  Helium would likely decrease the conductivity as it is not ionised under the cool conditions of $\sim$20,000~K or less expected in the interiors of Uranus and Neptune;  however, the $\lambda$ values presented by \citet{French12} for a H/He-mixture along a Jupiter adiabat are still larger than the H$_2$O values under the same conditions. 
\\For the inner envelope, the thermal conductivity of H-C-N-O mixtures under planetary interior conditions is not well constrained by experimental data or simulations. According to \citet{French19}, the electronic contribution to $\lambda_\text{H2O}$ becomes comparable to the ionic for $T \gtrsim \SI{6e3}{K}$ and dominant for $T \gtrsim \SI{10e3}{K}$. In this regime, $\lambda$ would behave roughly the same as the electrical conductivity $\sigma$. \citet{Ravasio21} find that, once fully ionised, the electrical conductivity of NH$_3$ is significantly higher than that of H$_2$O, which suggests that a H$_2$O-NH$_3$-mixture would have a higher $\lambda$ than pure H$_2$O in this regime. It is less clear how the addition of hydrocarbons would affect $\lambda$. \Citet{Chau2011} find a $10^{-4}$--0.1-times lower electrical conductivity in CH$_4$ than H$_2$O at lower pressures of up to 30 GPa. However, the thermal conductivity under the conditions of their experiments might contain significant contributions from the nuclei, so $\sigma$ in the experiments of \citet{Chau2011} might not be indicative of $\lambda$. Additionally, X-ray diffraction analysis of compressed hydrocarbons at higher pressures of 150~GPa and about 5000~K as typical for the deeper interior of the ice giants reveal phase separation of carbon from hydrogen and nano-diamond formation \citep{Kraus17}. In a real planet, diamonds may sink downward, and, by convective motion between the TBL region and the diamond rain region, deplete the TBL region from carbon. Therefore, the ice-rich region below the TBL may be better described by an ammonia and water mixture. Thus, in light of these uncertainties,  we used $\lambda_\text{H2O}$ as a lower bound.
In Sect. \ref{sec:res_conduct}, we discuss the influence of higher thermal conductivities on our models by introducing a factor between 5 and 1000 by which $\lambda_\text{H2O}$ is multiplied. Our discussions underline the need for reliable material data for complex H-C-N-O mixtures at extreme conditions.
\section{Results}
\label{sec:results}
\subsection{Influence of a TBL on evolution behaviour}
\label{sec:res_general}
\begin{figure}
  \includegraphics[width=0.46\textwidth]{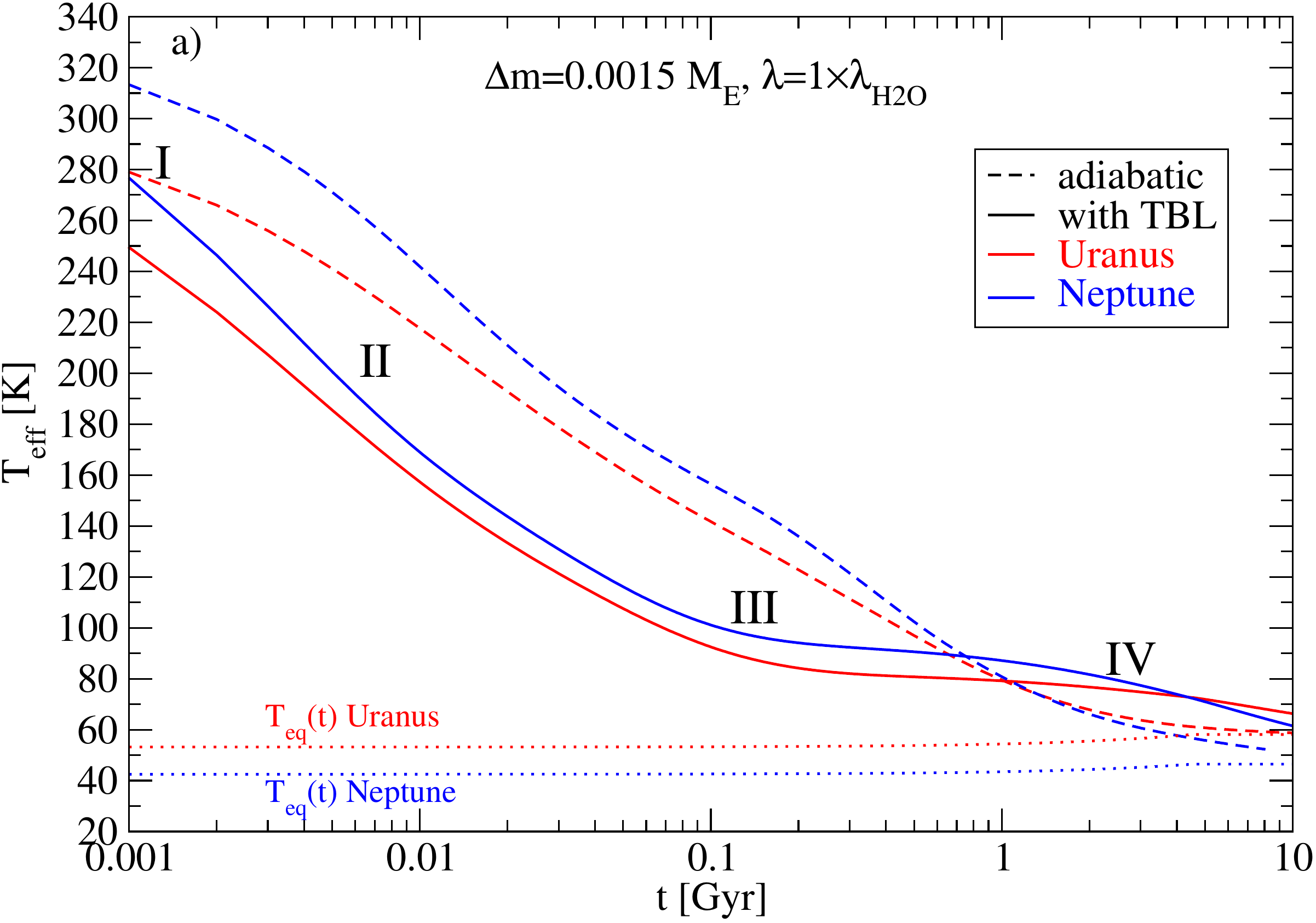}
  \includegraphics[width=0.49\textwidth]{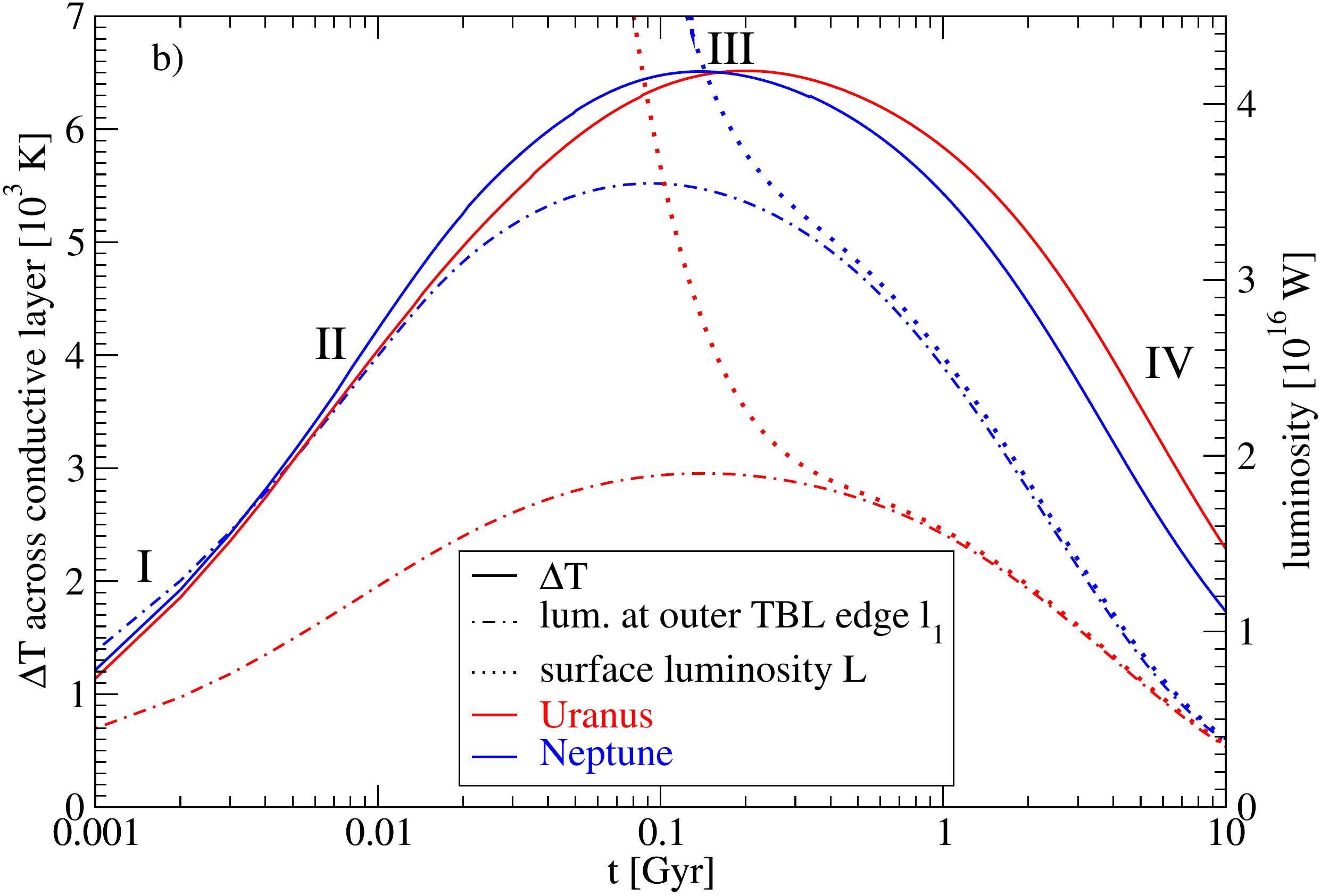}
  \caption{Evolution of (a) effective temperature $T_\text{eff}$ and (b) temperature difference across TBL $\Delta T$, luminosity at the outer edge of the TBL $l_1$, and planet surface luminosity $L$ with time for a Uranus-sized (red) and a Neptune-sized (blue) ice giant with a thin conductive interface of $\Delta m = 0.0015~M_\text{E}$. Dashed lines in (a) are adiabatic models of the same composition and dotted lines give the equilibrium temperatures (see text for discussion).}
  \label{fig:general}
\end{figure}
We outline typical behaviour of the models under the influence of a TBL using the example of a Uranus model and a Neptune model with a fixed TBL of $\Delta m = 0.0015~M_\text{E}$.  In the top panel of Fig.~\ref{fig:general}, we show the effective temperature $T_\text{eff}$ compared to an adiabatic model of equal metallicity. In the bottom panel, we show the temperature difference $\Delta T$ across the TBL as well as the local luminosity $l_{1}$ at the top of the TBL. We can distinguish four characteristic epochs: (I) the planet with a TBL quickly becomes fainter compared to the adiabatic case. This is due to the fact that inefficient heat transport across the interface traps the bulk of the planet's energy in the deep interior, leaving mainly the outer envelope to cool. Our models give a large $\Delta T \sim 1000$~K already after the first time step, probably due to the hot start and the sudden switch from a completely adiabatic interior to a conducting interface. During phase II, $\Delta T$ across the TBL rises continuously as essentially only the outer envelope cools. Because a steeper temperature gradient enables a higher conductive heat flux to be transported, the luminosity at the outer edge of the TBL rises steadily.
During this time, the outer envelope radiates heat away and contracts, so the planet's surface luminosity $L$ decreases (see dotted lines in Fig. \ref{fig:general}b). Since this contraction deposits more gravitational energy in the interior than can be radiated away through the TBL, the inner envelope can even slightly heat up over the first 100 Myr.\\
Towards the end of phase II, the temperature at the top of the TBL is significantly lower than in the beginning (low $T(m)$), the cooling of the atmosphere is less efficient (lower $T_\text{eff}$  and lower $L$), and further contraction of the initially fluffy, gaseous envelope is countered by repulsive particle interaction as the matter becomes denser. As the rising $l_1$ and the falling $L$ approach each other (see Figure \ref{fig:general}b), the decrease in $L$ prevents $l_1$ from rising further, since the former acts as a cap to the latter. Consequently, $l_1$ adopts a maximum, which marks the beginning of phase III.\\
In phase II and into phase III, the outer envelope and atmosphere are significantly cooler and more contracted than in an adiabatic planet. However, the inner envelope constitutes an enormous energy reservoir that keeps feeding the outer envelope and slows its further cooling: $T_\text{eff}$ can adopt a plateau-like behaviour. Since the planet is still cooling and both $L$  and $R$ are decreasing, albeit slowly, $l_1$ is forced to do the same. Because it is predominantly the inner envelope that cools during phase III, $\Delta T$ across the TBL now also begins to decrease. Phase III can last for billions of years.\\
After some time has elapsed, the bulk of the energy of the deep interior has been released and the cooling begins to accelerate again (phase IV). Notably, in phase IV the planet can appear brighter than an adiabatically cooled planet of the same age. \citet{Leconte13} showed that Saturn's high luminosity can be explained by a very similar behaviour, caused in their model by a region of layered convection. 
The behaviour outlined above applies to our models of thin to moderate TBL thicknesses and all considered  $\lambda$ values.\\
\citet{Vazan20} estimated the minimum thickness for a conductive layer to meaningfully affect the planetary evolution of an ice giant to be about $\Delta r > \SI{100}{km}$, which is more than the TBL thicknesses considered in this work. They used the diffusion timescale for thermal conduction $\tau_c=\Delta r^2 \rho c_p/\lambda$  and requested a timescale of $1\text{ -- }5$~Gyr. Using our values of $\rho=\SI{0.3}{\gram\per\cubic\centi\metre}$, $\lambda = \SI{2}{\watt\per\metre\per\kelvin}$, $c_p=\SI{15}{\kilo\joule\per\kilogram\kelvin}$, and TBL thicknesses in the range $\Delta m = 0.0015\text{ -- }0.03~M_\text{E}$ ($\Delta r = 4\text{ -- }50~\si{km}$), we obtain shorter timescales of $\tau_\text{cond} \sim 1\text{ -- }100~\si{Myr}$. As shown in Figure 4, we also find that $\Delta T$ across the TBL increases to several 1000~K within only a few million years, and that this early change in the cooling behaviour has a significant influence on the long-term evolution. Therefore, our thinner TBLs compared to the estimate by \citet{Vazan20} are consistent with the short timescales influencing the evolution seen in our results.
%
%
\subsection{Conducting interface thickness $\Delta m$}
\label{sec:res_thick}
\begin{figure}
  \includegraphics[width=0.49\textwidth]{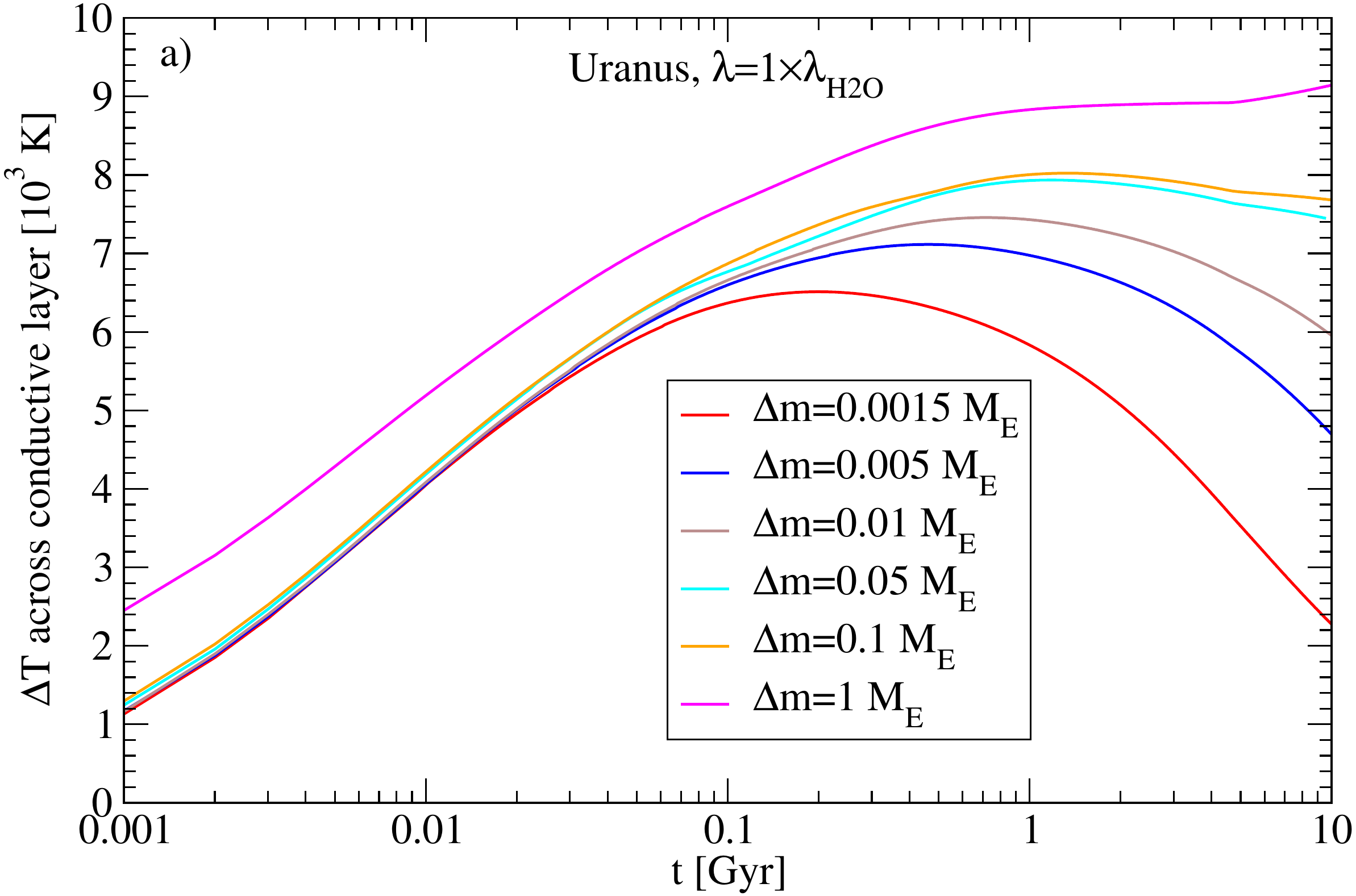}
  \includegraphics[width=0.49\textwidth]{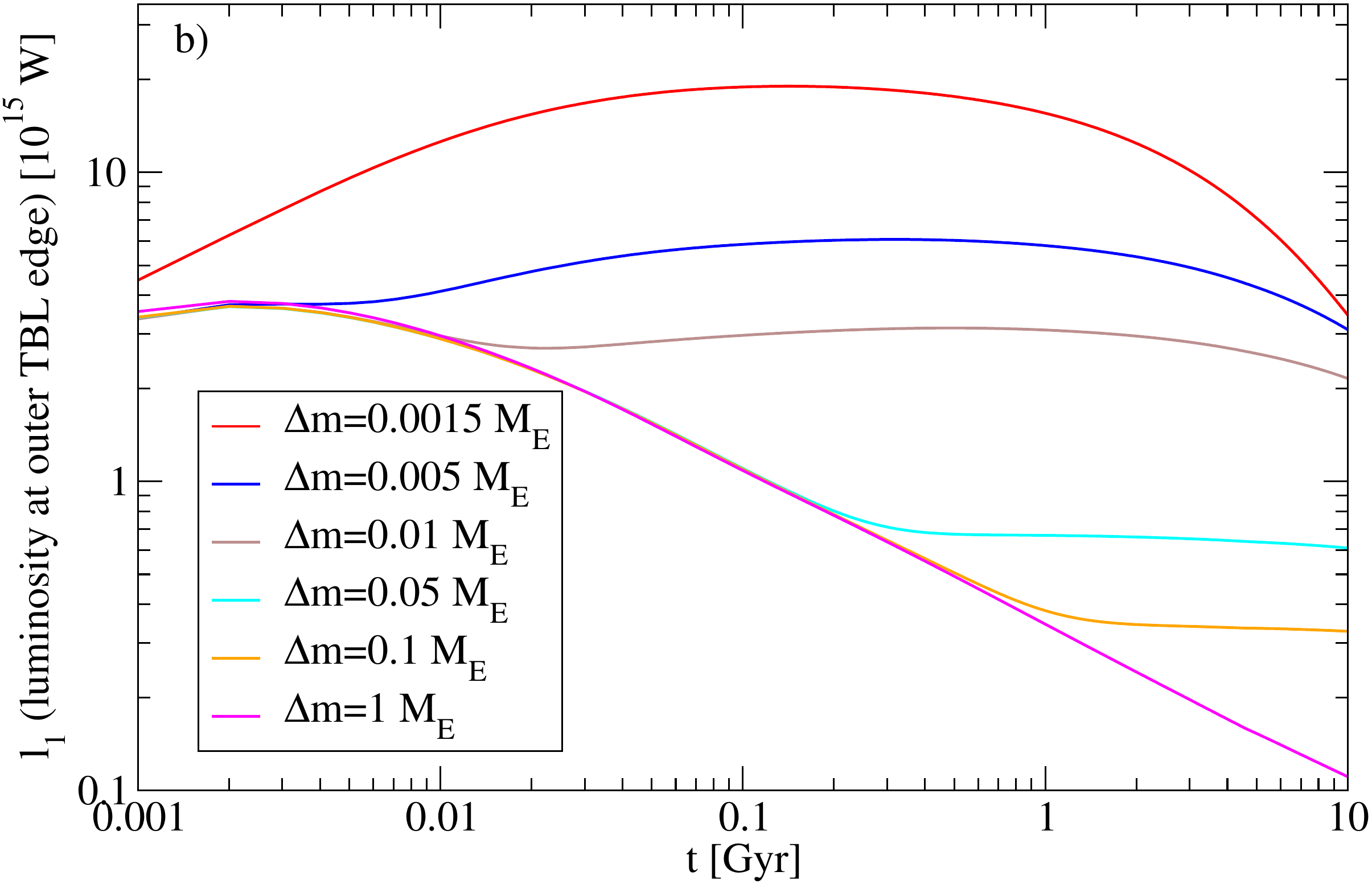}
  \includegraphics[width=0.49\textwidth]{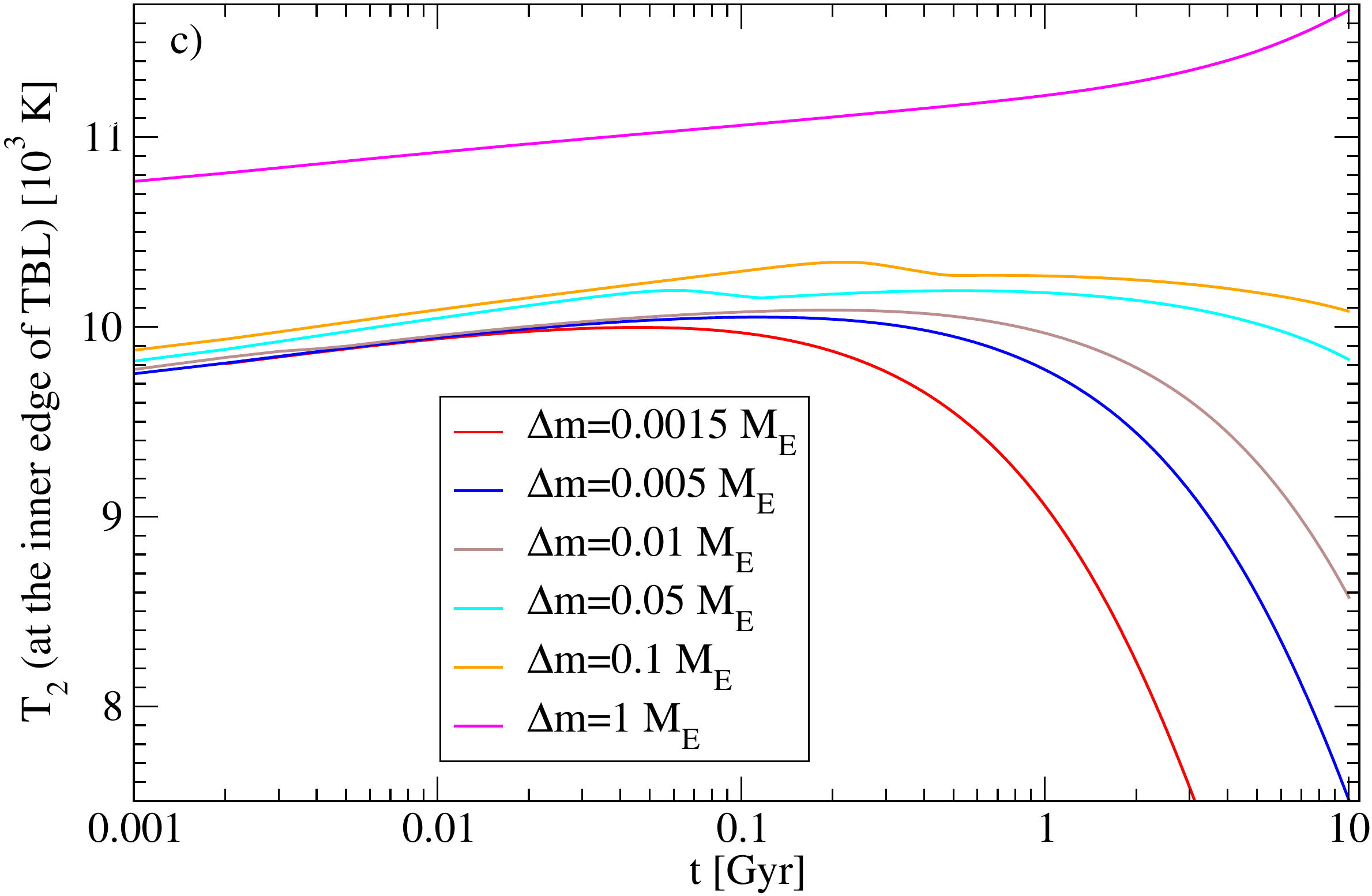}
  \caption{Evolution of TBL for a Uranus model of varying TBL thickness $\Delta m$ and   $\lambda=\lambda_\text{H2O}$. Top: Temperature difference $\Delta T$ across TBL; middle: luminosity $l_1$ at the outer edge of the TBL; bottom: temperature $T_2$ at the inner edge of the TBL.} 
  \label{fig:VaryDmDT}
\end{figure}

\begin{figure}
  \includegraphics[width=0.49\textwidth]{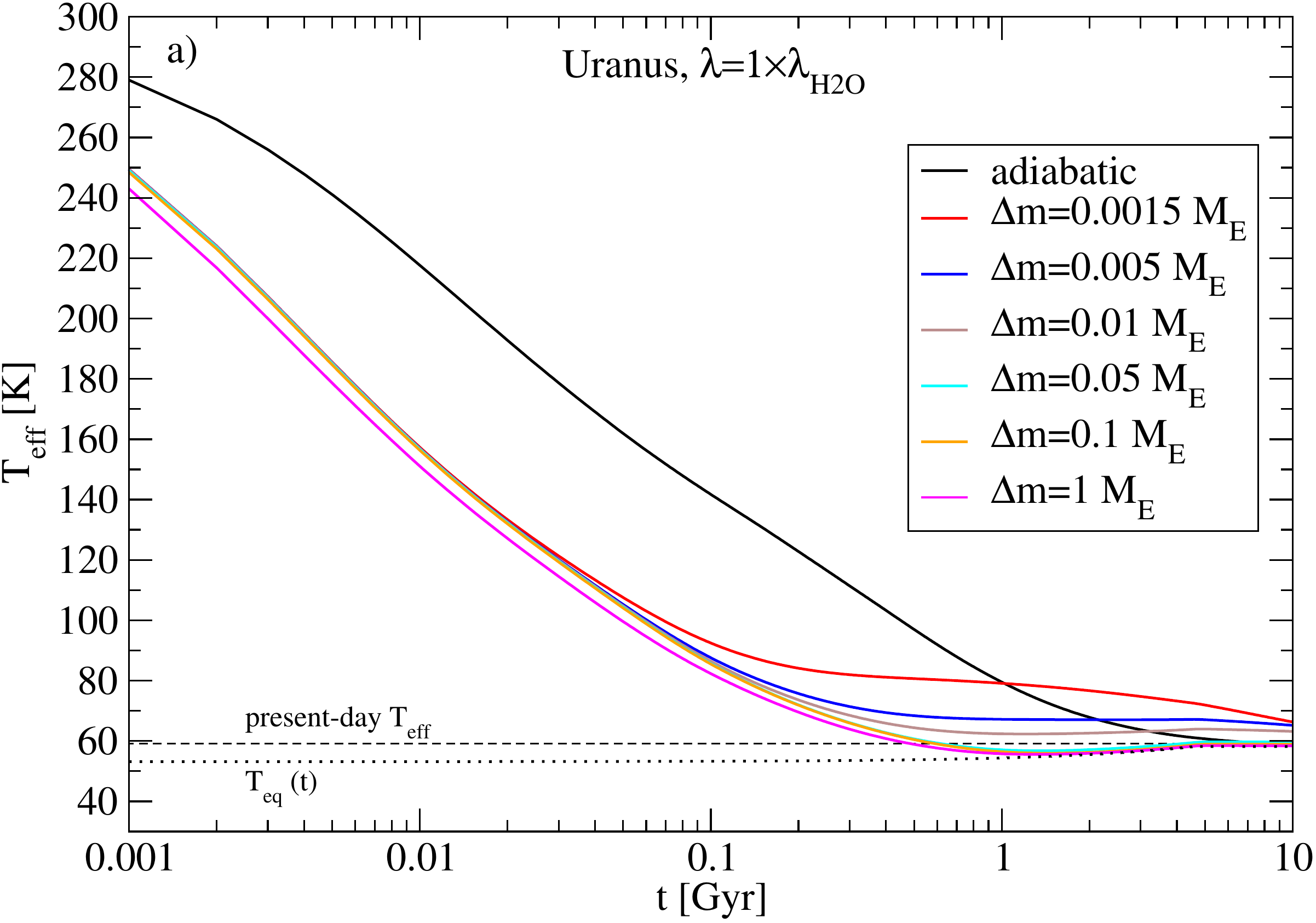}
  \includegraphics[width=0.49\textwidth]{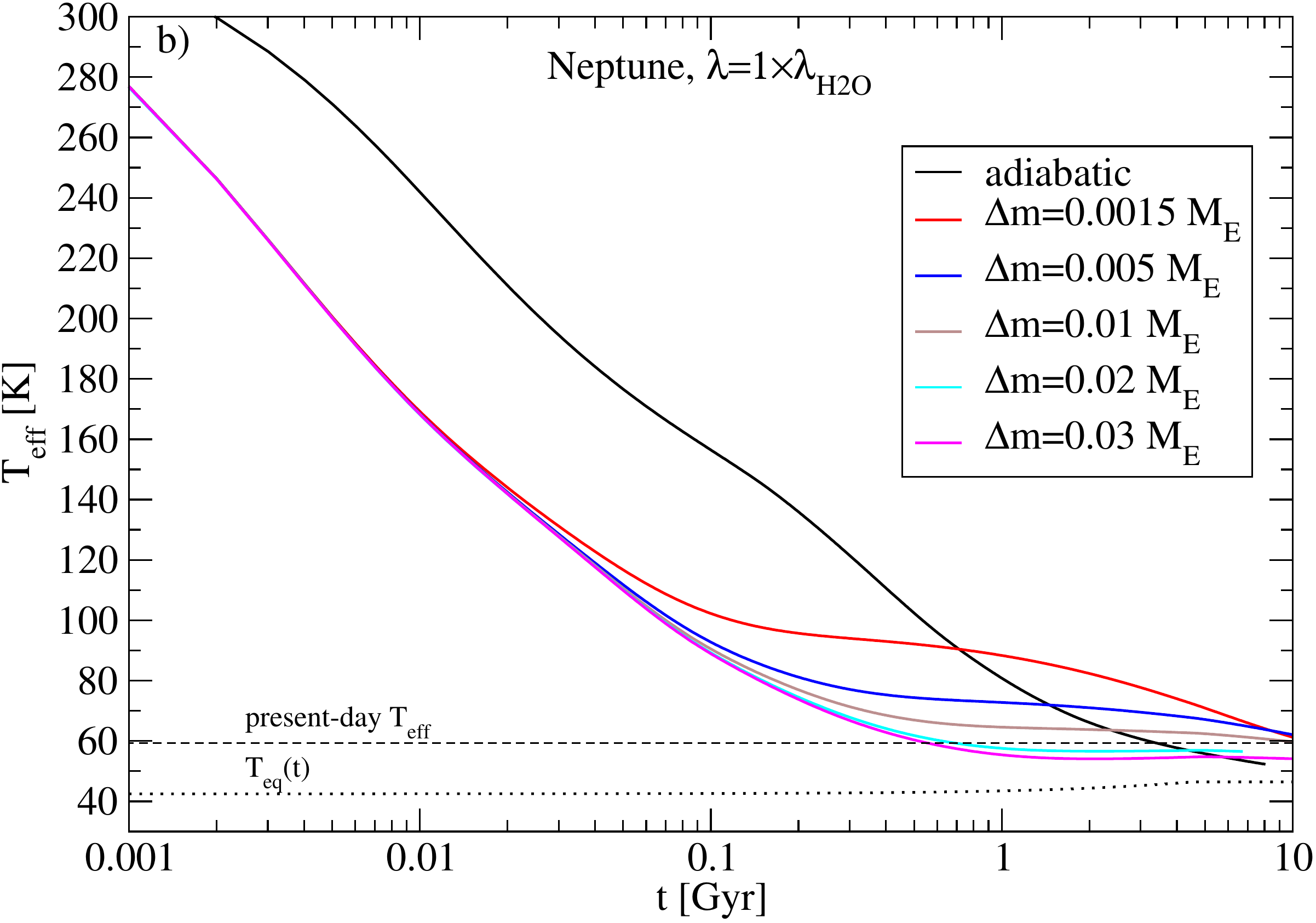}
  \caption{Thermal evolution of a Uranus model (a) and a Neptune model (b) with a conductive TBL of varying thickness (coloured curves) between the outer and inner envelope compared to an evolution model of the same composition, but without TBL (black curves). Dotted line: $T_\text{eq}(t)$; dashed line: present-day $T_\text{eff.}$} \label{fig:VaryDm}
\end{figure}

We investigated the influence of different thicknesses of the assumed conductive layer on the planetary evolution. In Fig.~\ref{fig:VaryDmDT}, we show $\Delta T(t)$ across the TBL, the luminosity $l_1(t)$ at the outer edge of the TBL, and the temperature $T_2(t)$ at the inner edge of the TBL for a Uranus model with different layer thicknesses $\Delta m$. Figure~\ref{fig:VaryDm} shows $T_\text{eff}(t)$ of a Uranus model and a Neptune model, as well as an adiabatic comparison case of equal composition.
A thicker TBL leads to less efficient energy transfer across it. Therefore, the luminosity $l_1$ decreases with increasing $\Delta m$ (Fig.~\ref{fig:VaryDmDT}b). Conversely, to transport the same amount of energy, a thicker conductive layer requires a stronger temperature incline and it takes longer to reach these steeper gradients (Fig.~\ref{fig:VaryDmDT}a). This leads to phase III happening later in the planet's lifetime for larger $\Delta m$. By that time, the outer envelope is even  cooler and more contracted than for small $\Delta m$, which causes phase III to feature a flatter plateau in $T_\text{eff}$ (see Figure \ref{fig:VaryDm}).  \\
If the preceding phase II of accelerated cooling of the outer shell lasts long enough, $T_\text{eff}$ at the transition to phase III can be so low that it approaches the equilibrium temperature $T_\text{eq}$, although complete equilibrium is never reached. This makes it almost impossible for the outer envelope to radiate away heat, which means the planet cannot cool and its temperature profile stays largely the same over a long time period. When $T_\text{eff}$ follows $T_\text{eq}$ (see Fig.~\ref{fig:VaryDm}a for Uranus), even a slight re-heating of the entire planet due to insulation is possible (both $\Delta T$ and $T_2$ rise). Such  models have hot deep interiors. Equilibrium evolution happens for Uranus over a wide parameter range. For Neptune, we see the plateau-like behaviour towards higher $\Delta m$, and since its $T_\text{eq}$ is significantly lower than that of Uranus, we do not observe this re-heating in our considered models (Fig.~\ref{fig:VaryDm}b).
\subsection{Influence of thermal conductivity $\lambda$}
\label{sec:res_conduct}
\begin{figure}
  \includegraphics[width=0.49\textwidth]{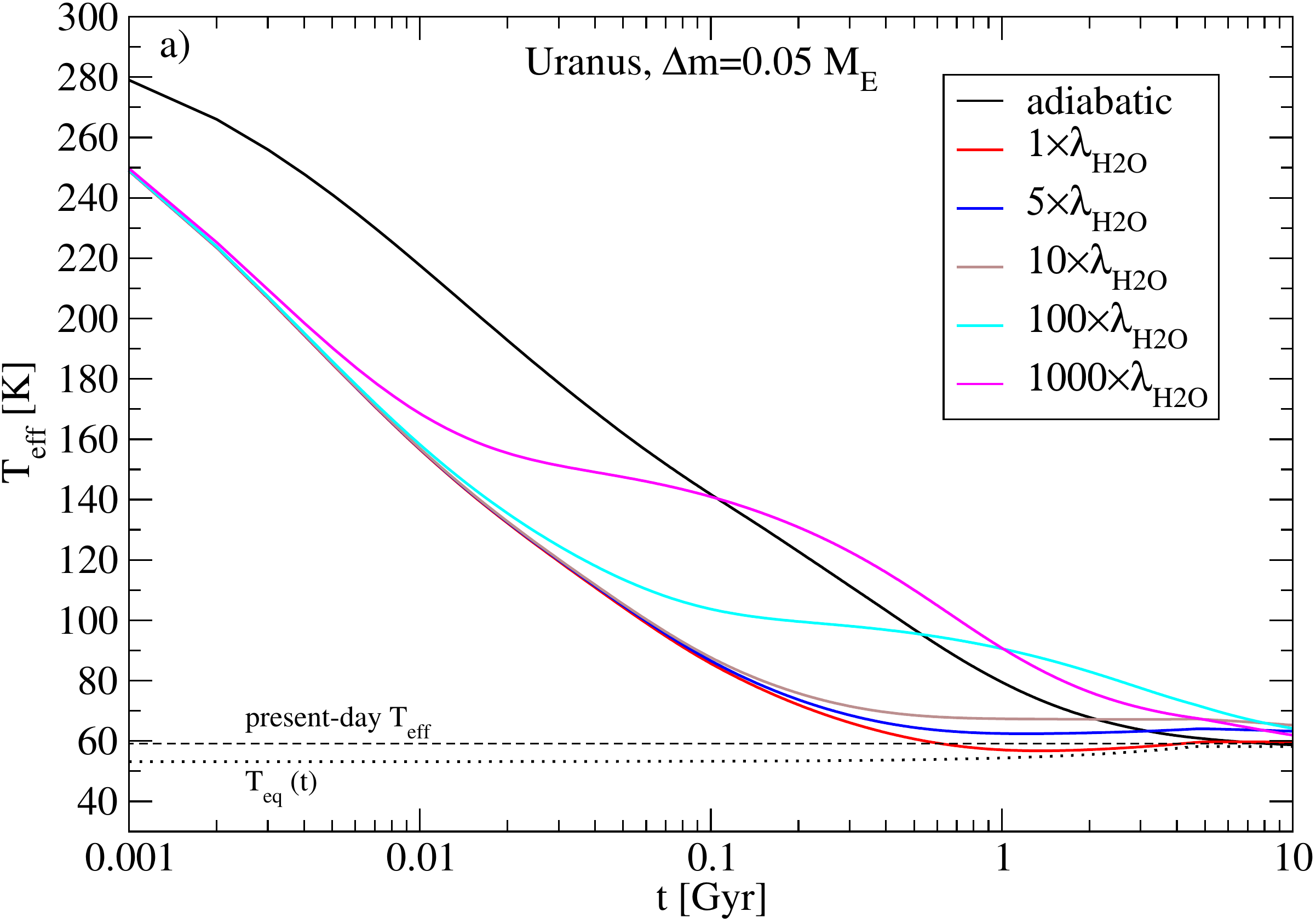}
  \includegraphics[width=0.49\textwidth]{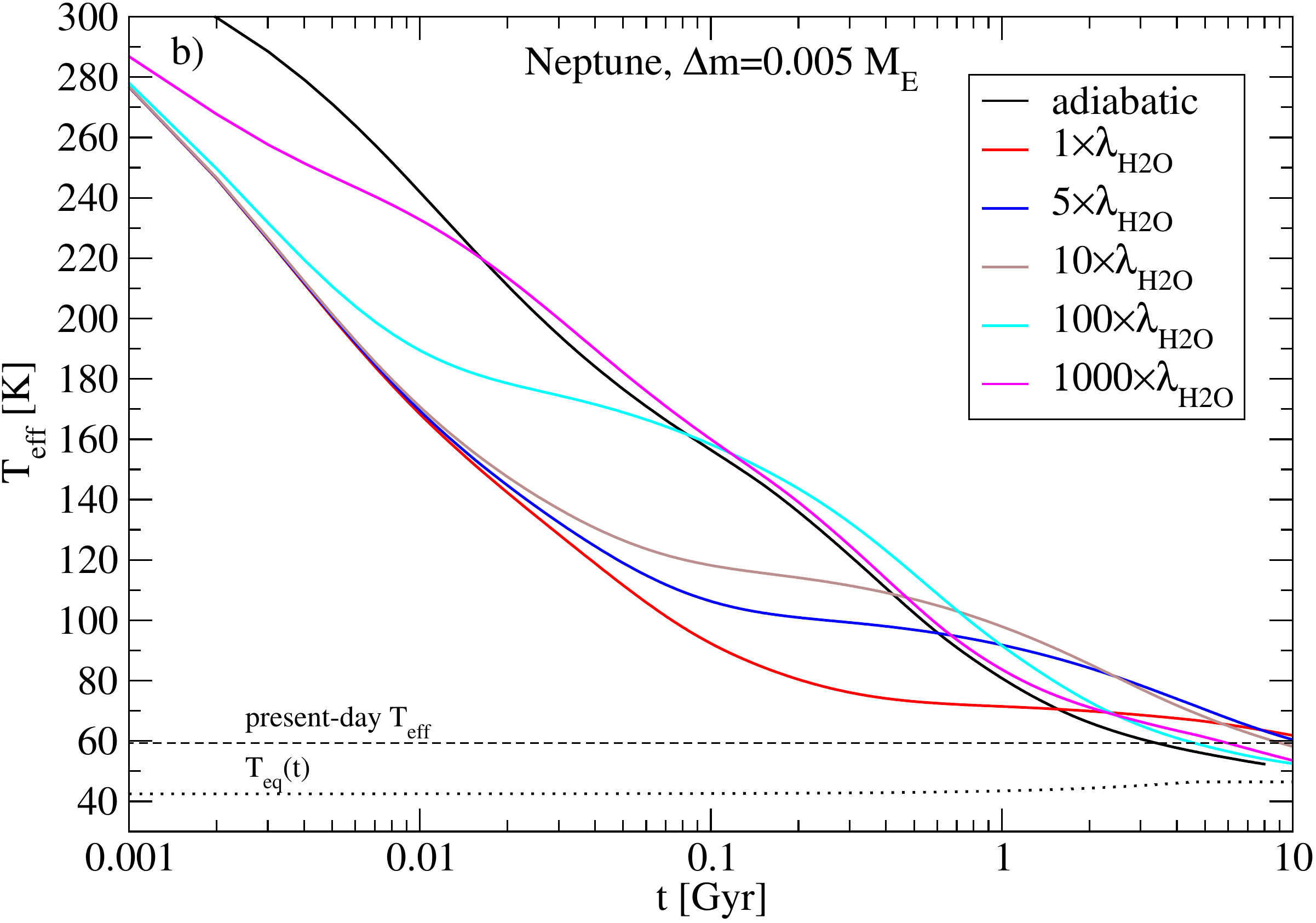}
  \caption{Influence of thermal conductivity value on the thermal evolution of a Uranus model (a) and Neptune model (b) with fixed TBL thickness $\Delta m$.} \label{fig:VaryLambda}
\end{figure}

As discussed in Sect.~\ref{sec:met_conduct}, using values of $\lambda$ for pure H$_2$O is a simplification for the complex mixture of materials forming the TBL and is taken to be a lower bound. Additionally, taking the TBL to be a purely conducting region is a further simplification. There might be energy transport processes that are more efficient than pure heat conduction, such as semi-convection. Since both of these factors can lead to more potent energy transport across the TBL than assumed in our baseline models, we investigated this possibility by calculating models with enhanced values for the thermal conductivity on our models. Figure \ref{fig:VaryLambda} shows evolutionary tracks for a conductive layer of fixed $\Delta m=0.05$~$M_\text{E}$ for Uranus and $\Delta m=0.005$~$M_\text{E}$ for Neptune and pure water in the inner mantle. Enhanced conductivity values were achieved by multiplying the \textit{\emph{ab initio}} water values by factors of $5$, $10$, $100,$ and $1000$ (cf. Sect.~\ref{sec:met_conduct}). $\lambda$ values enhanced by $10^2$ or $10^3$ are highly unlikely to occur for the purely conductive case, because, for example, even a pure hydrogen plasma features $\lambda < \SI{e3}{\watt\per\kelvin\per\metre}$ for conditions comparable to our models \citep{Holst11}. These higher factors are included primarily to simulate more effective transport processes across the TBL. \\
The general behaviour remains similar to that of the original case ($1\times \lambda_\text{H2O}$, cf. Sect.~\ref{sec:res_general}); we still clearly see an accelerated decrease in $T_\text{eff}$ compared to the adiabatic case (phase II), followed by slower cooling (phase III), followed again by another acceleration (phase IV). However, for higher $\lambda$ values the transition from evolution phase II to phase III occurs at earlier times and thus at higher temperatures. This is because energy transport through the conductive layer is more efficient for higher $\lambda$ values. Heat from the deep interior can contribute to the overall evolution more efficiently already in phase II. This way, the energy from the inner envelope reaches the outer envelope at a time when it is still extended and thus better able to radiate away energy. As a result, the slowdown of the heat loss during phase III is weaker for higher $\lambda$ values. In the late stages of  the evolution, the curves converge towards the adiabatic case, except for $\lambda=1000\times \lambda_\text{H$_2$O}$, where the conductive gradient in the whole inner envelope becomes smaller than the adiabatic one, thus making the inner envelope wholly conductive, which visibly influences the cooling behaviour. The results from Sects.~3.2 and 3.3 show that enhancing $\lambda$ acts in the opposite direction to enhancing $\Delta m$. 
\subsection{Interior profiles}
\label{sec:res_profiles}
\begin{figure}
  \includegraphics[width=0.49\textwidth]{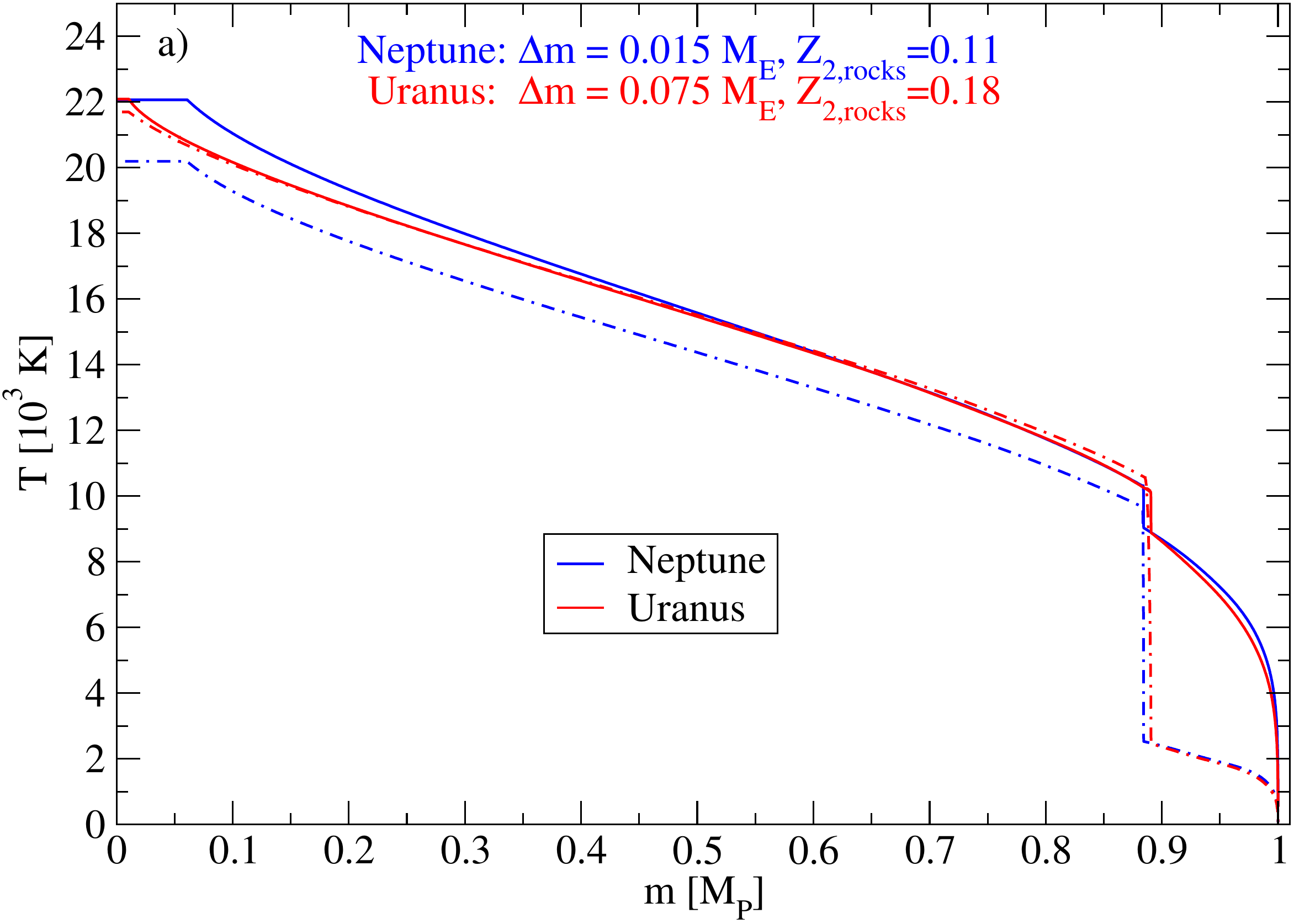}
  \includegraphics[width=0.49\textwidth]{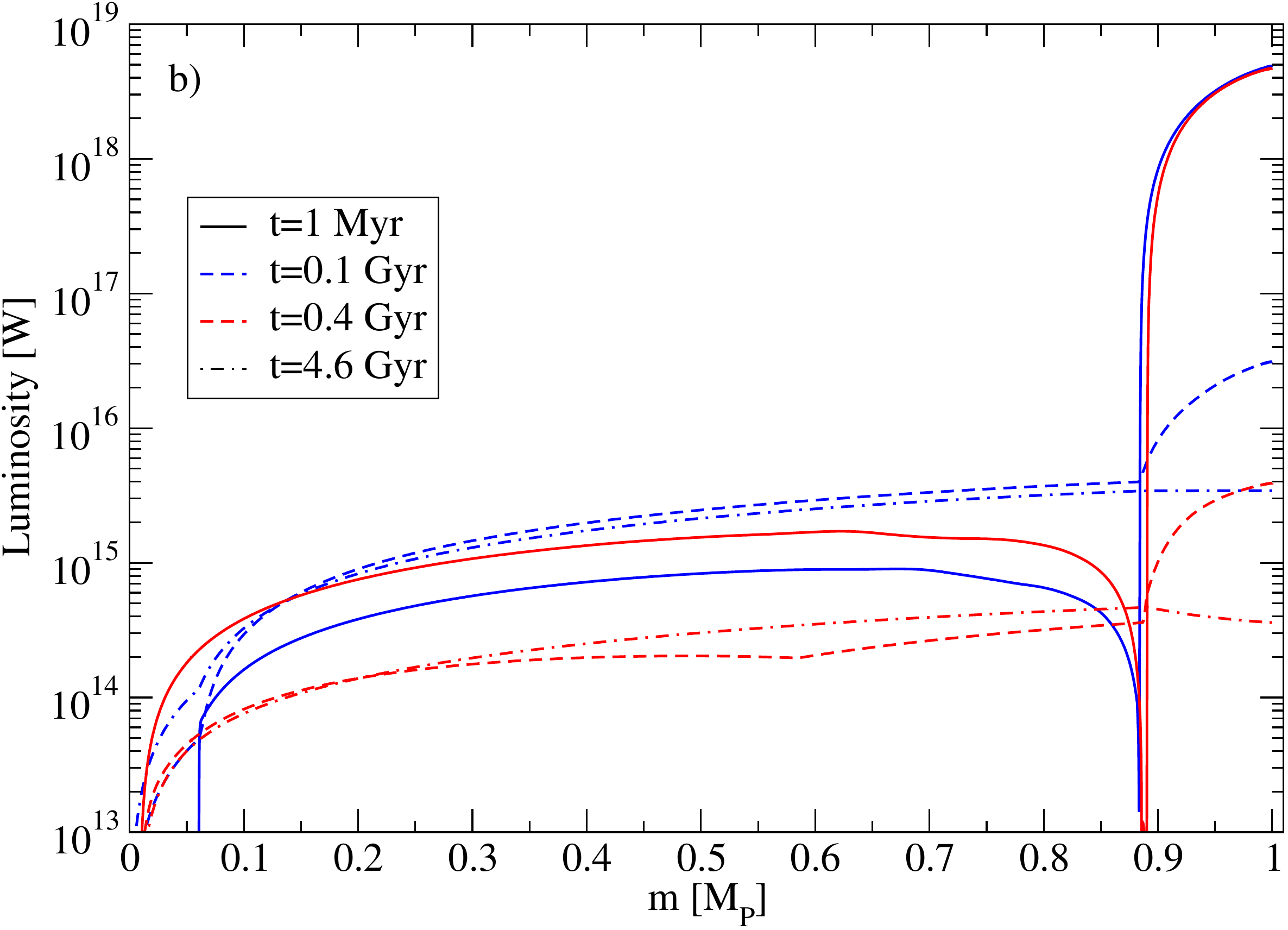}\\
  \caption{Interior profiles for Uranus (red) and Neptune (blue) models at different time steps that reproduce today's radius and effective temperature.} \label{fig:InnerProfiles}
\end{figure}
\begin{figure}
  \includegraphics[width=0.49\textwidth]{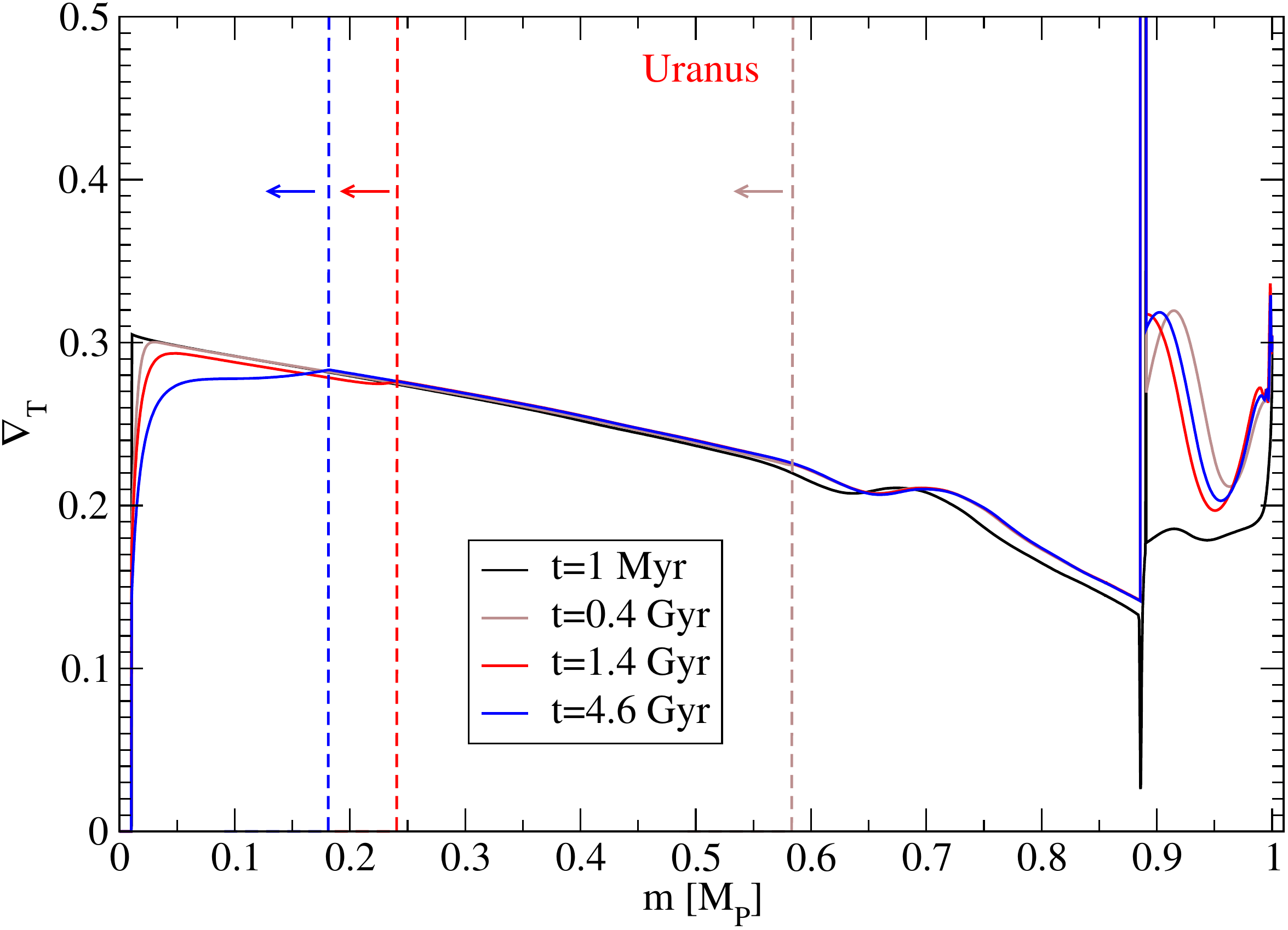}
  \includegraphics[width=0.49\textwidth]{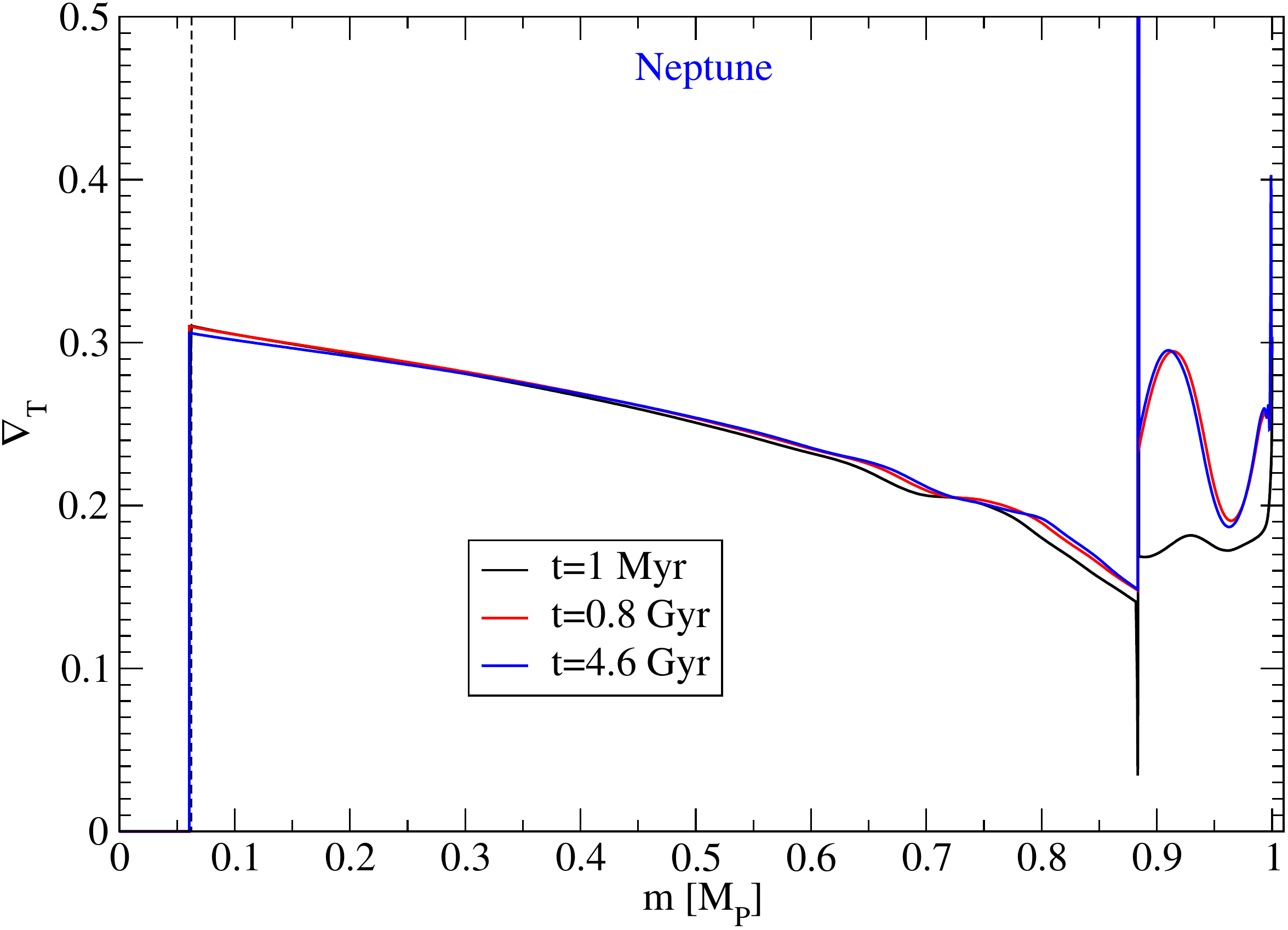}\\
  \caption{Temperature gradient profiles for the same Uranus (a) and Neptune (b) models as in Figure \ref{fig:InnerProfiles}. The value at the outside of the conductive interface is higher than the maximum displayed $\nabla_T$ value by several orders of magnitude (cf.\ Fig.\ \ref{fig:GradCompare}). The vertical dashed lines mark the upper end of the conductive zone in the deep interior.} \label{fig:GradTProfiles}
\end{figure}
In this section, we discuss profiles of models that,  by the choice of $\Delta m$, $\lambda$, and $Z_{2,\,\rm rocks}$, reproduce today's observed effective temperature and radius. The Uranus model has $\Delta m=0.075$~$M_\text{E}$, $1\times\lambda$, and $Z_{2,\,\rm rocks}=0.18$, while the Neptune model has $\Delta m=0.015$~$M_\text{E}$, $1\times\lambda$, and $Z_{2,\,\rm rocks}=0.11$. Figure \ref{fig:InnerProfiles} shows the profiles of $T$ and $l$ along the mass coordinate at the beginning of the evolution and at present time at $t=\SI{4.6}{Gyr}$. For early times, the local luminosity $l$ shows a strong dip at the location of the conductive interface. While the outer envelope can cool efficiently and thus the surface luminosity $L$ is high, $l$ goes close to zero at the TBL, indicating low net energy transport to the outside. For some models, this dip can even reach small negative numbers during phases I and II, where energy is actually transported to the inside since contraction and cooling of the outer envelope increases the pressure and deposits gravitational energy. As can be seen in Fig.\ \ref{fig:VaryDmDT}c, this leads to a slight heating of the inner envelope compared to the initial model. As time passes, the luminosity dip shrinks until it has completely vanished. For Uranus, by the time phase III starts, the outer envelope has already reached a state close to equilibrium with the solar incident flux, which means efficient cooling is no longer feasible. The inner envelope's temperature remains close to the hot starting model and the luminosity becomes low throughout the planet. In contrast, as Neptune is farther away from the Sun, it can still cool after several $\si{Gyr}$ and thus has higher luminosity in keeping with the observations. However, even though the Neptune model allows for some cooling of the deep interior, there is still a strong temperature increase of $\sim \SI{7000}{K}$ across the TBL at $t=\SI{4.6}{Gyr}$.\\
Figure\ \ref{fig:GradTProfiles} shows the temperature gradient within the planets. For Uranus we find that, early on, the inner envelope features several regions that are stably stratified to convection according to the Schwarzschild criterion. This is due to the low local luminosity and thus low conductive temperature gradient (cf.\ Eq.\ \eqref{eq:nabla_con}). These regions shrink, and after about $t=\SI{0.4}{Gyr}$  we mostly see a single deep conductive and stable region extending upward from the core. The stable area shrinks steadily from the top over the planet's remaining lifetime and is still present at $t=\SI{4.6}{Gyr}$, although reduced to within $\sim20\%$ of the Uranus mass. Comparing this to the Neptune model, which has a smaller interface, we find no similar region developing beyond the first few time steps, after which the whole inner envelope remains convective. This is likely due to the fact that the smaller interface does not isolate this interior as effectively, meaning the luminosity is greater in these models. We find comparable behaviour for Uranus models with the same smaller interface thickness. %
\subsection{Present effective temperature}
\label{sec:res_Teff}
\begin{figure}
  \includegraphics[width=0.49\textwidth]{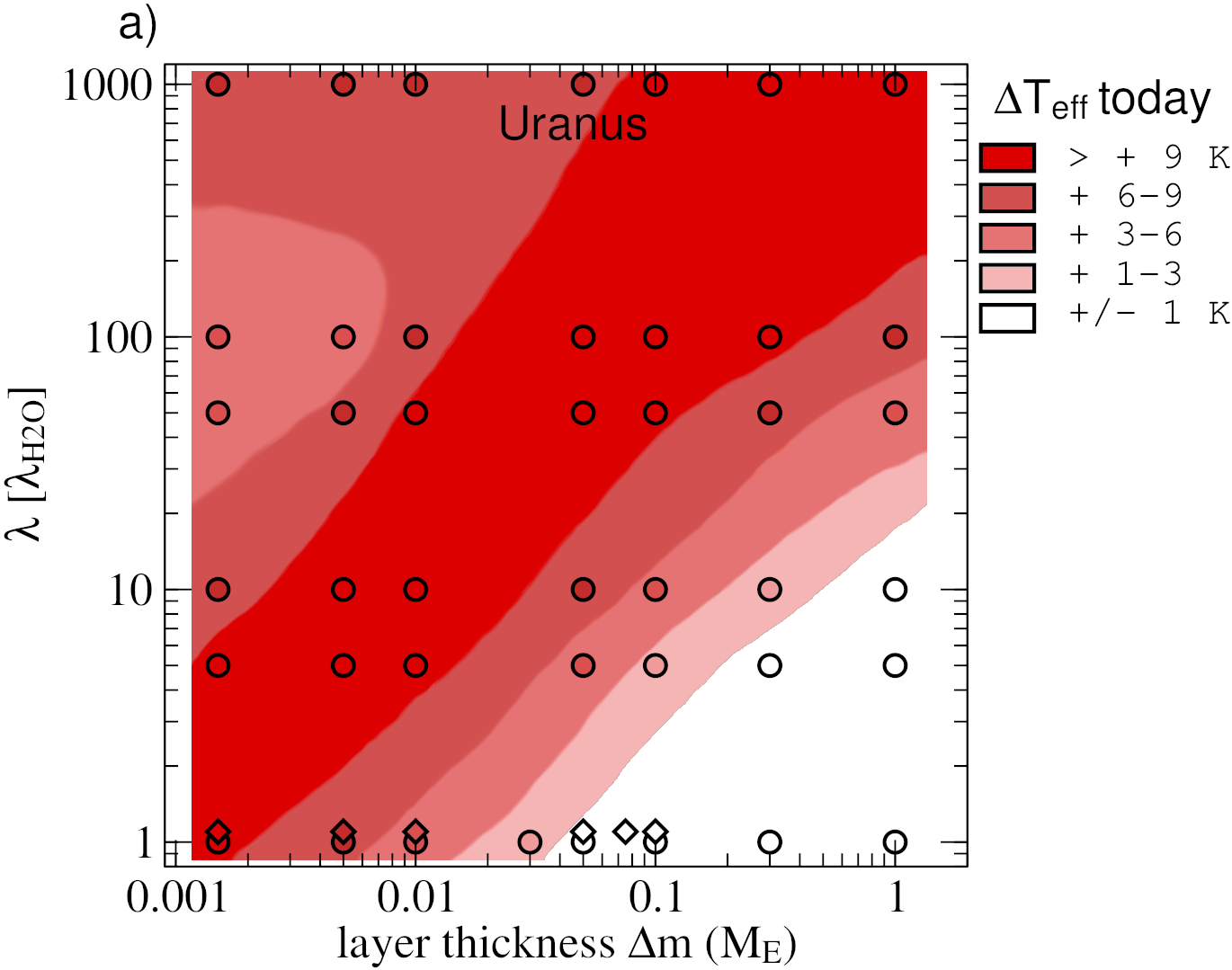}
  \includegraphics[width=0.49\textwidth]{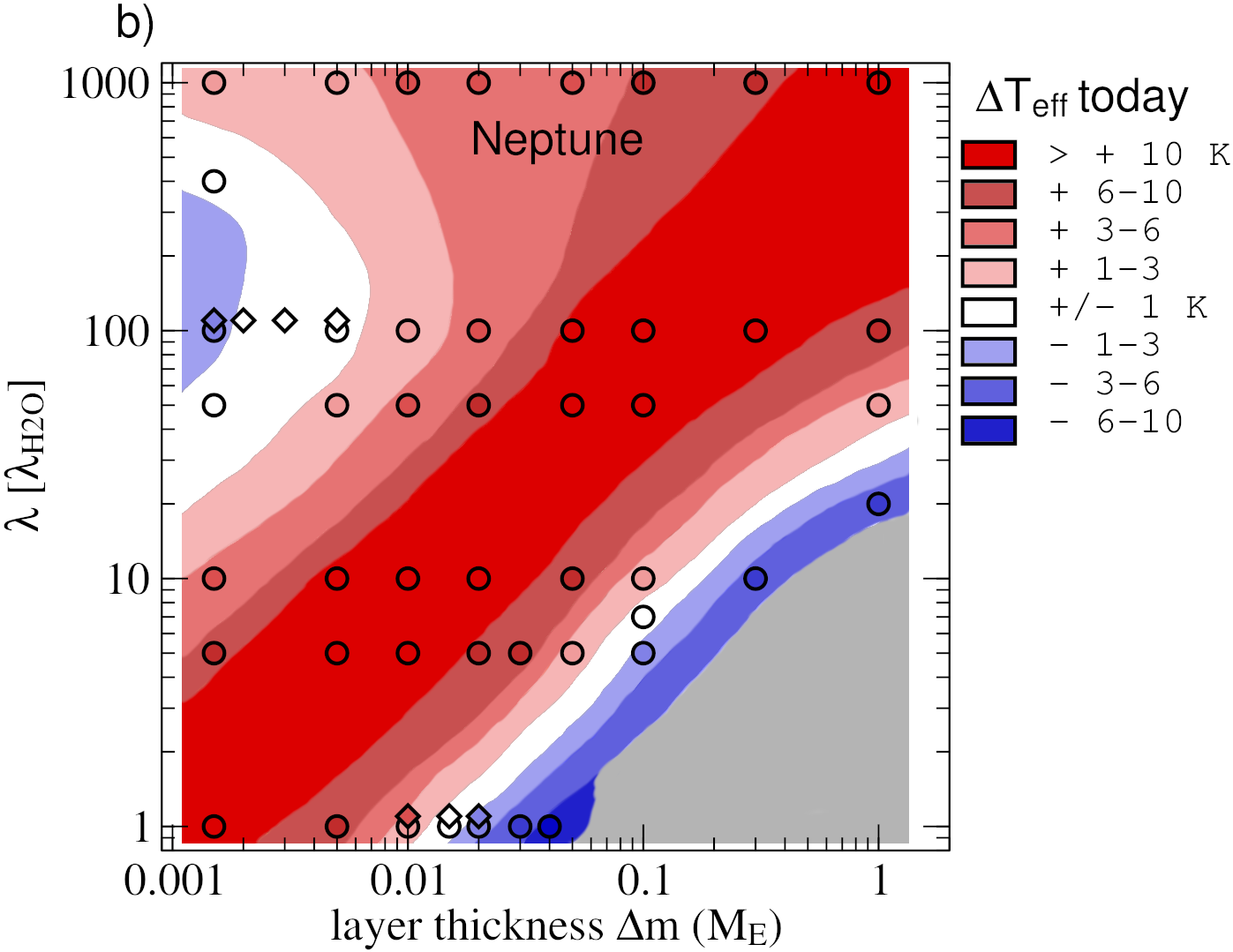}
\caption{Global view on present-day effective temperatures for a number of Uranus (a) and Neptune (b) evolution models of different TBL thicknesses $\Delta m$ and thermal conductivity factors. Models with $T_\text{eff}\sim$ observed $T_\text{eff}$ are shown in white, those in red tones are too hot, and those in blue tones are too cold. The models employ a composition of $Z_{2,\text{water}}=1$ in the inner mantle, except those shown as diamonds -- near the panel bottoms and in the white halo in b) -- which are calibrated to reproduce the observed $R_\text{P}$. The background colour serves as rough guide to the eye. The grey area marks space where no data is available.}
\label{fig:scatter}
\end{figure}
In this section, we differentiate between models that satisfy the observed luminosity and those that do not. In Figure\ \ref{fig:scatter}, we present $T_\text{eff}$ at $t=\SI{4.56}{Gyr}$ as a function of assumed layer thickness and thermal conductivity factor. Models with a $T_\text{eff}$ close to the observations are shown in white. Those that are too hot are shown in red tones, and those in blue tones are too cold. For both planets there are models for moderate layer thicknesses in the 0.02--0.05$~M_\text{E}$ ranges ($\sim$ 30 to $\sim$\SI{100}{km}) that agree with the observed $T_\text{eff}$. In fact, for Uranus, all the models with $\Delta m > 0.02~M_\text{E}$ are close to today's luminosity. Comparing to, for example, Figure\ \ref{fig:VaryDm}, we see that this is due to the fact that for these values Uranus evolves close to the equilibrium temperature over long timescales. For Neptune on the other hand, which differs from Uranus mainly in that its $T_\text{eq}$ is noticably smaller than its  $T_\text{eff}$, we find two disjunct sets of models that are valid in this sense. One occurs for moderate $\Delta m$ and $\lambda=1\times \lambda_\text{H2O}$. In this case the planet resides in phase III with a $T_\text{eff}$-plateau near the observed $T_\text{eff}$. The other case occurs for thin interfaces and $\lambda\sim 100\times\lambda_\text{H2O}$. Here, a completed phase III has prolonged Neptune's cooling just enough (compared to the adiabatic evolution) that it hits the right $T_\text{eff}$ today. In this second set of solutions, the TBL delays the release of energy only over a short ($\sim 10$--$\SI{100}{Myr}$) period of time, before the planet resumes cooling behaviour similar to an adiabatic one. This suggests that even a transient phase of reduced convective or semi-convective cooling could sufficiently delay planetary cooling for Neptune to reproduce the observed luminosity. However, this class of solutions would predict Uranus to be $\sim$3--6~K hotter than observed, while the class of solution with formation of a plateau in $T_\text{eff}$ yields viable models for both planets. \\
We also see a pattern of diagonal structures in Figure\ \ref{fig:scatter}, where an increase in $\Delta m$ can be compensated for by a higher $\lambda$-factor to still give the same present-day result. This makes sense since both a thicker interface and lower conductivity make for less efficient energy transport across the boundary layer. \\
All models depicted as circles in Fig.\ \ref{fig:scatter} use $\SI{100}{\%}$ H$_2$O in the inner envelope. Since the temperature of the deep interior varies from model to model, this also leads to a variation of present-day radius, which means that almost none of the models presented in Fig.\ \ref{fig:scatter} reproduce today's observed mean radius. For $\lambda=\lambda_\text{H2O,}$ adding about $\SI{10}{\%}-\SI{20}{\%}$ of the rocks to the inner envelope is required to achieve the observed radius, depending on the interface thickness, which in turn leads to a variation of up to $\SI{2}{\kelvin}$ in the present-day $T_\text{eff}$. For an enhanced conductivity $\lambda/\lambda_\text{H2O}\sim 100$, about $\SI{7}{\%}$ of H and He have to be added to the inner envelope. The diamond-shaped points in Fig.\ \ref{fig:scatter} show solutions for which this adjustment has been made.
They reproduce the observed radius to a deviation of less than \SI{0.3}{\%}. Thus, while the models in Figure\ \ref{fig:scatter} do not represent actual Uranus and Neptune by radius within the observational error bars, the described trends nevertheless hold true. Specifically, the general regions where we expect solutions that match $T_\text{eff, obs.}$ only change by $\sim 2$~K due to this simplification. Adjusting the rock content of every model to fit today's observed radius proved impractical and time consuming due to the high computation time involved in the calculations. 
\subsection{Boundary layer stability}
\label{sec:res_stability}
\begin{figure}
  \includegraphics[width=0.49\textwidth]{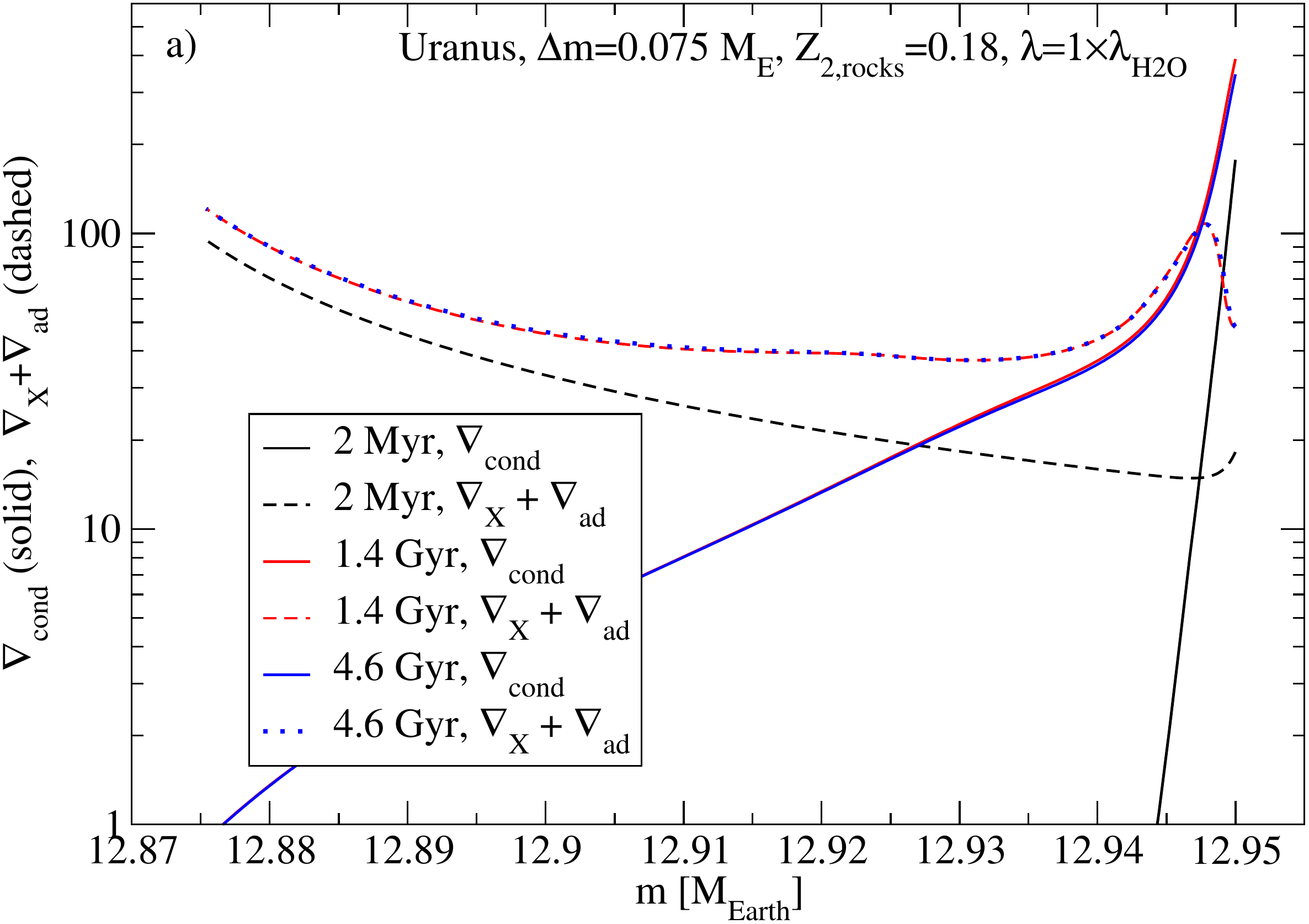}
  \includegraphics[width=0.49\textwidth]{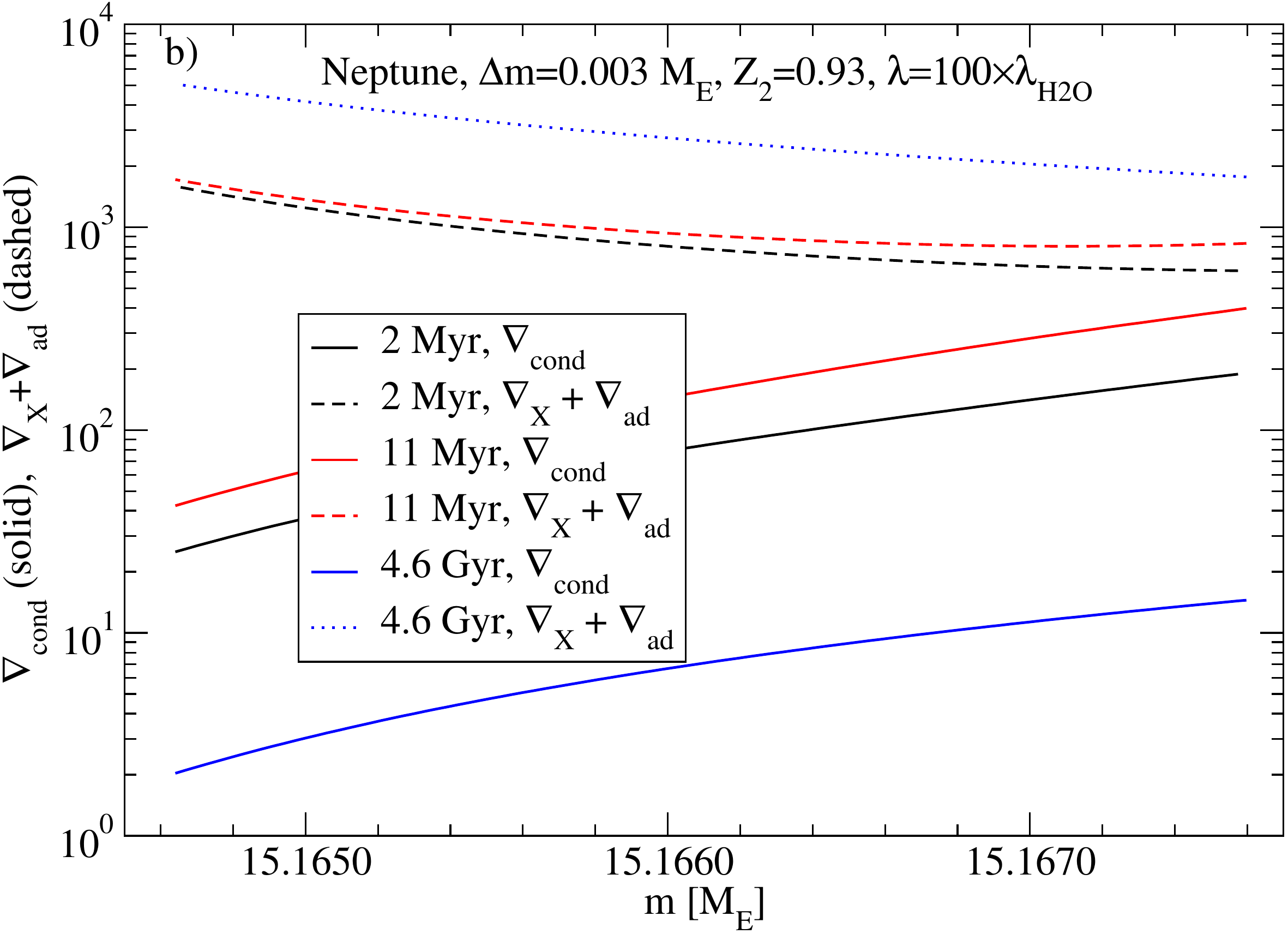}
  \caption{Gradients in Eq.\ \eqref{eq:LedouxX} across the TBL for a Uranus model with $\Delta m = 0.075 M_\text{E}$, $\lambda=\lambda_\text{H2O}$ (top panel) and a Neptune model with $\Delta m = 0.003 M_\text{E}$, $\lambda=100\times\lambda_\text{H2O}$ (bottom panel). Chosen times are the second time step, the time of maximum $\Delta T$ across the TBL, and the present time, respectively. Regions where $\nabla_{\text{cond}}$ exceeds $\nabla_X + \nabla_{\text{ad}}$ would be unstable to convection.} \label{fig:GradCompare}
\end{figure}
To assess the stability of the thin conducting interface, we used the Ledoux criterion formulated in terms of mass fractions \citep{Vazan15}:
\begin{align}
    \nabla_{\text{cond}} - \nabla_{\text{ad}} \leq \nabla_X , \label{eq:LedouxX}
\end{align}
with  $\nabla_{\text{cond}}, \nabla_{\text{ad}}$ from Eqs. \eqref{eq:nabla}, and 
\begin{align}
  \nabla_X= - \frac{1}{T}\sum_i  \derive{X_i}{\ln P}  \dfrac{ \left( \partdiff{\rho}{X_i} \right)_{P, T, X_{m\neq i}} } { \left( \partdiff{\rho}{T} \right)_{P, X}  }.
\end{align}
Regions that fulfil the inequality \eqref{eq:LedouxX} are stably stratified against convection. \\
In Figure\ \ref{fig:GradCompare}, we show $\nabla_{\text{cond}}$ and $\nabla_{\rm ad}+\nabla_{X}$ across the TBL for a Uranus evolution
model (a) and a Neptune evolution model (b) that both fulfil the observed $T_\text{eff}$ and $R_\text{p}$. For Uranus, we see that throughout the evolution, the temperature gradient in the very outer part of the TBL exceeds the stabilising compositional gradient that would make that particular region unstable to convection. In the majority of the conducting layer, however, the compositional gradient is strongly dominant, thus stabilising the region. The instability at the top of the layer might lead to partial erosion of the TBL and eventually cause the compositional profile to assume a step-like structure similar to that found by \cite{Vazan18}. Considering convective mixing and the dynamical growth and shrinking of stable layers based on the Ledoux criterion is beyond the scope of this work.\\
The Neptune model for which we address boundary layer stability is from the white halo of solutions in Fig.~\ref{fig:scatter} and features both a thinner TBL and a significantly higher $\lambda$ value than the Uranus model. Since this means both a steeper compositional gradient (cf. Fig. \ref{fig:Zm}) and a lower temperature gradient, the interface remains stable against convection over the entire evolution period.\\ 
That the conductive temperature gradient varies so strongly in this relatively thin part of the planet can be attributed to two factors: $\nabla_{\text{cond}}$ is directly proportional to the local luminosity $l,$ and it is inversely proportional to the temperature $T$. Within the boundary layer, the temperature rises sharply over several thousand Kelvins (see Fig.\ \ref{fig:VaryDmDT}), which naturally leads to $\nabla_{\text{cond}}$ falling. For early times, this is reinforced by the fact that $l$ transitions from a high value in the outer envelope, where energy can flow relatively unimpeded, to a much lower one in the inner envelope where heat flux is very low. \\
\subsection{Delayed TBL onset}
\label{sec:res_late}
\begin{figure}
  \includegraphics[width=0.49\textwidth]{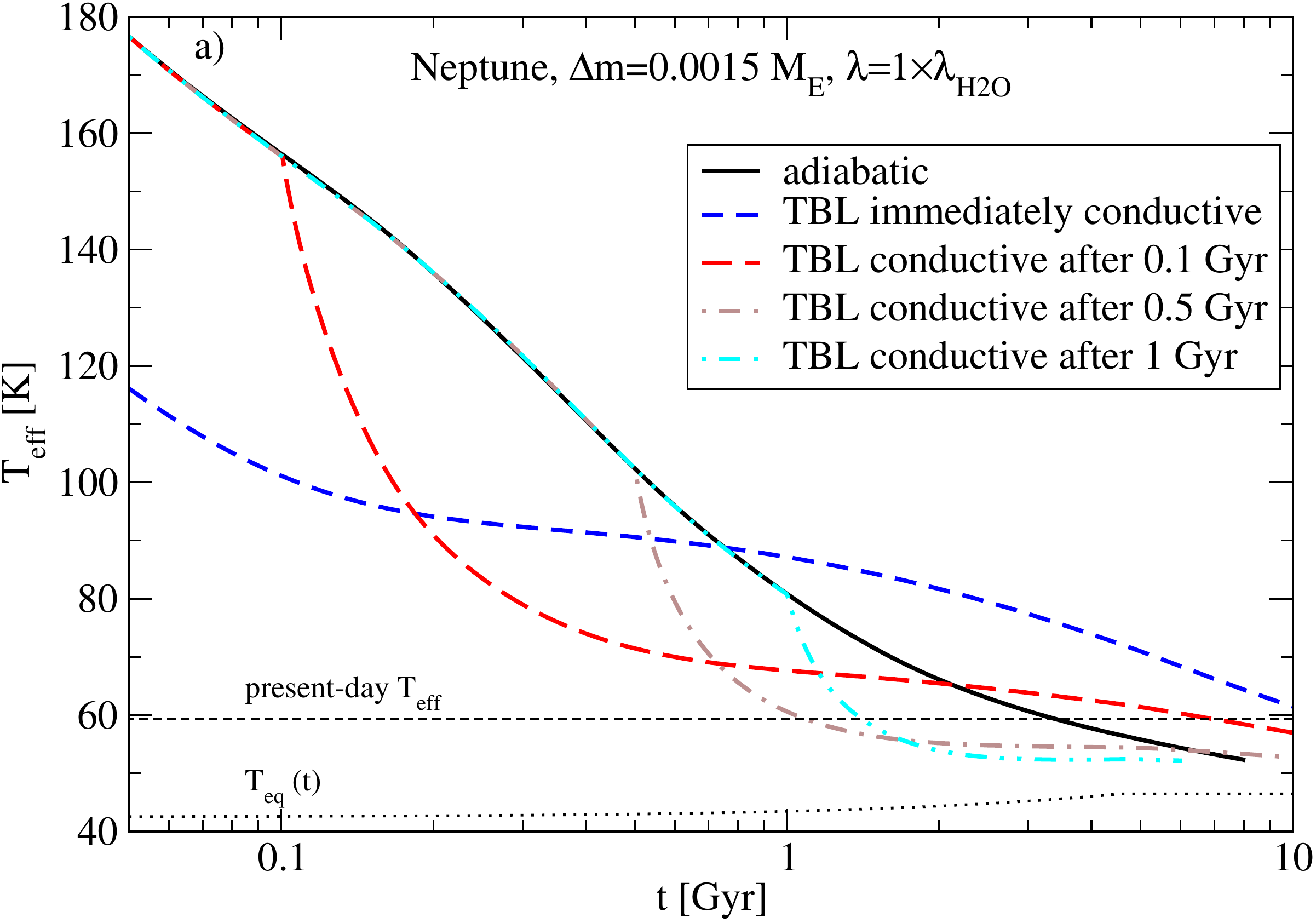}
  \includegraphics[width=0.49\textwidth]{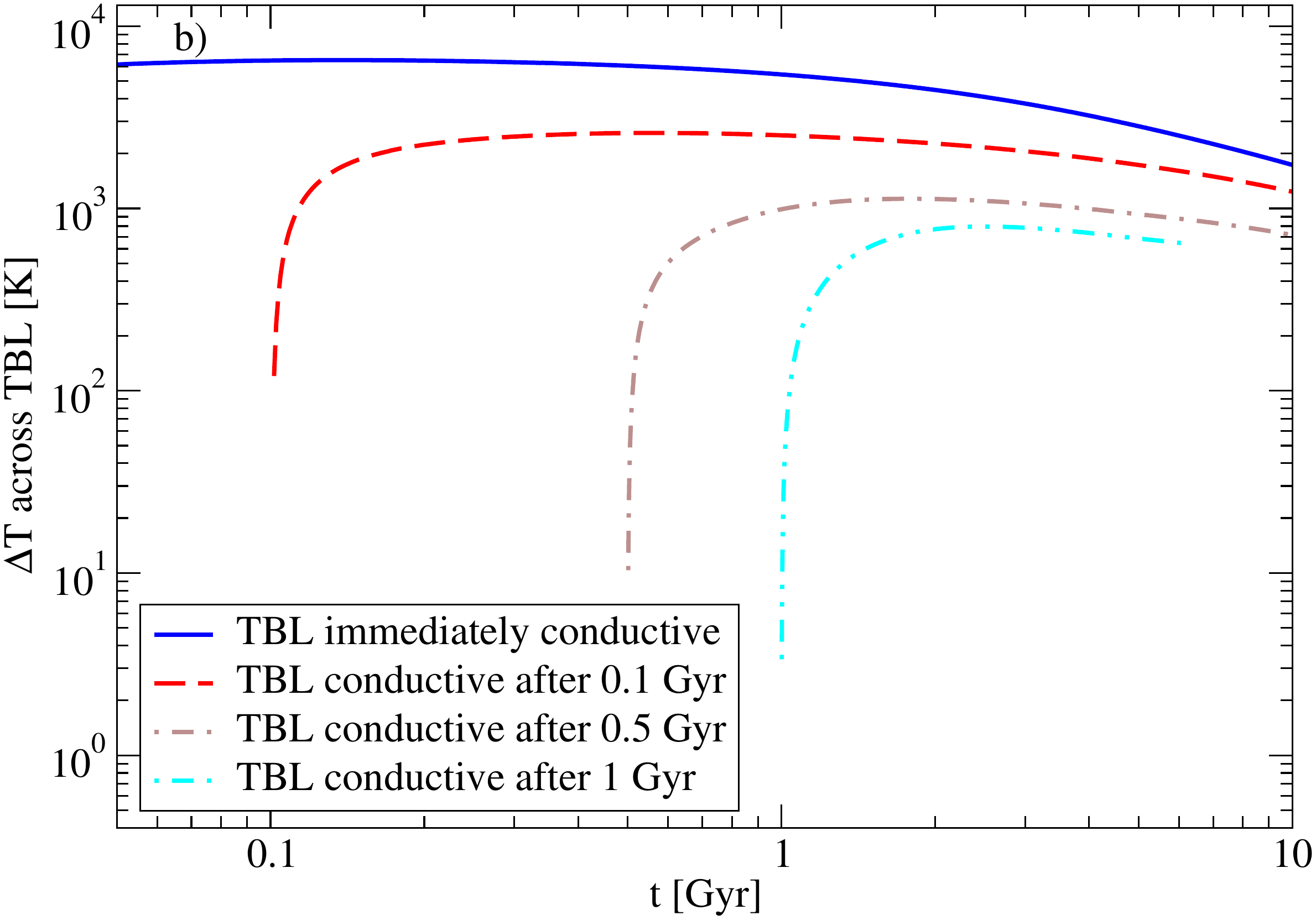}
  \caption{Evolution of a Neptune model and a Uranus model with $\Delta m = 0.0015~M_\text{E}$ for different times after which the TBL is switched to conductive. a: Effective temperature, dotted line: $T_\text{eq}(t)$, thin dashed line: present-day $T_\text{eff}$; b: temperature difference across TBL.}
  \label{fig:lateN}
\end{figure}
\begin{figure}
  \includegraphics[width=0.49\textwidth]{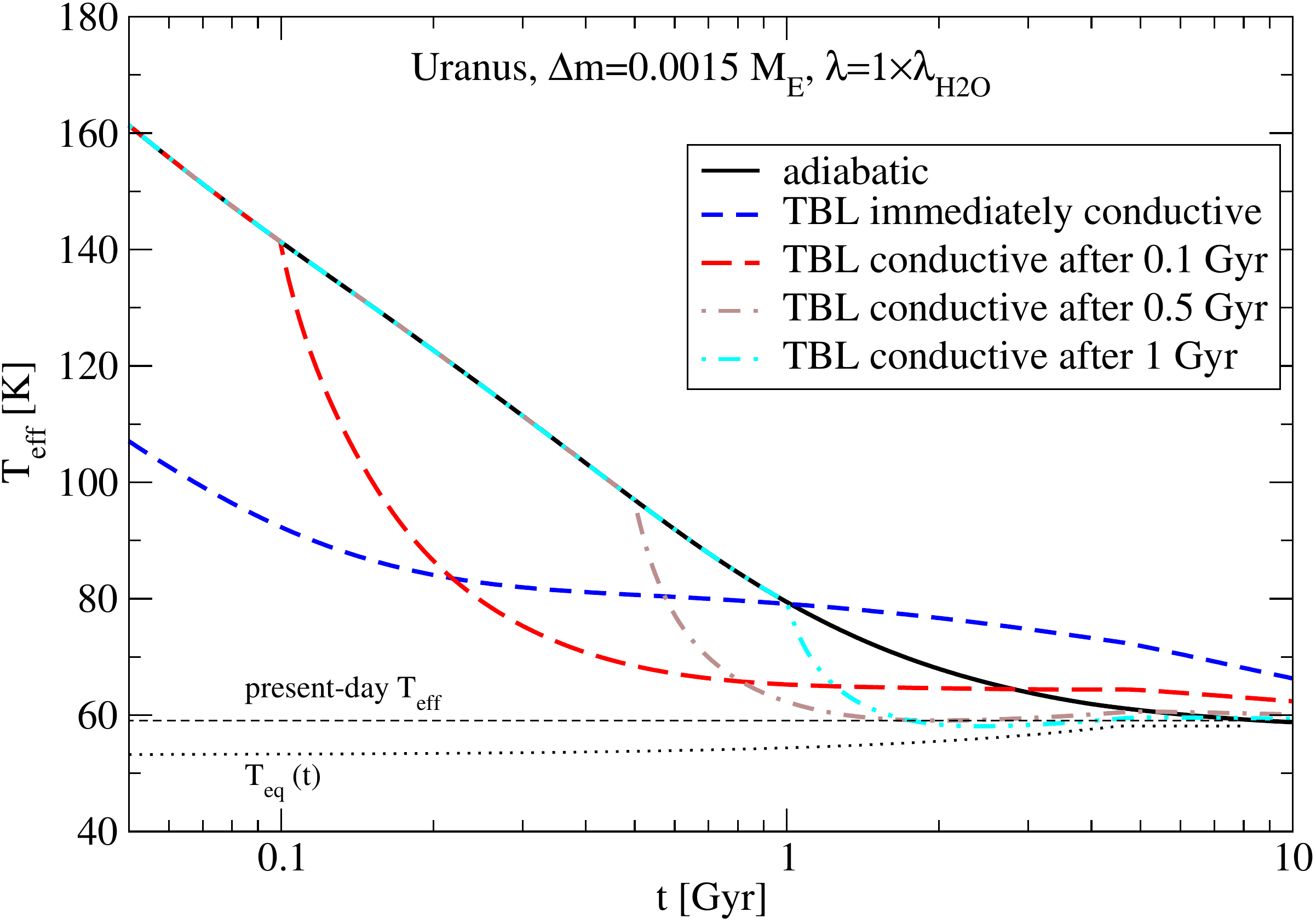}
  \caption{Same as Fig. \ref{fig:lateN} a), but for Uranus.}
  \label{fig:lateU}
\end{figure}
The models presented so far assume the presence of a conductive TBL from the onset of the planetary evolution. However, this does not necessarily need to be the case. The compositional differentiation that leads to the conductive layer could be the result of a number of physical processes; for example, the demixing of the ices from hydrogen and helium \citep{Bailey21}, which only occurs at lower temperatures and thus later in the planetary lifetime. To get an understanding of how the time after which such a TBL develops affects the evolution, we investigated a simple test case: we assumed all layers, including the interface between outer and inner envelope, to be adiabatic up to a certain point in time. After that point is reached, the boundary layer is switched from convective to conductive energy transport. Figure\ \ref{fig:lateN} shows the evolution of such a Neptune model with $\Delta m = 0.0015~M_\text{E}$. Once the TBL is switched on, the cooling of the exterior is accelerated, corresponding to phases I and II in Sect.~\ref{sec:res_general}. However, the initial $\Delta T$ across the TBL can be an order of magnitude lower now because at this point in the evolution, energy flux in the planet is substantially lower than in the beginning, and thus so is $\nabla_\text{cond}$. The later the conductive energy transport is enforced, the lower the maximum $\Delta T$ across the TBL becomes. 
As seen in Fig. \ref{fig:lateU}, we found similar behaviour in a Uranus model of the same $\Delta m$, except that, as discussed previously, Uranus models cannot go substantially below the observed $T_\text{eff}$ due to it being close to solar equilibrium. \\
The possibility of a delayed onset of a conductive interface broadens the parameter space for solutions that match the observed $T_\text{eff}$ considerably. For example, a very thin TBL with $\lambda=\lambda_\text{H2O}$ would otherwise leave both Uranus and Neptune too bright (cf.\ Fig.\ \ref{fig:VaryDm}), while a delayed onset could bring such models into agreement with the observed luminosity. The Uranus models are relatively insensitive to the assumed onset time as long as it occurs after $\sim\SI{0.5}{Gyr}$. This is in contrast with Neptune, where an earlier onset would be required to hit the observed $T_\text{eff}$. \\
It should be noted that these results merely give a basic guidance on how a delayed TBL by itself would influence a planet's cooling behaviour. If the differentiation and thus the TBL indeed formed via sedimentation, for example H$_2$-H$_2$O demixing, this would release additional gravitational energy, which in turn would raise the planet's overall luminosity, similarly to H$_2$-He separation inside Saturn, for example \citep{Stevenson77, Puestow16, Mankovich20}. For Uranus, this possible additional heating from sedimentation requires the presence of a counteracting process such as a TBL even more to delay the energy release. For Neptune, for which \citet{Bailey21} suggest ongoing rather than completed differentiation, this additional luminosity could conceivably lift some of our too faint solutions up to the observed brightness. Ultimately, the exact thermal effect of internal sedimentation and the resulting composition change remains to be investigated in future works, where the demixing of H$_2$ and H$_2$O has to be specifically included, a behaviour, that is at present unfortunately not well understood.
\section{Conclusions}
\label{sec:conclusions}
We present a range of evolution calculations for both Uranus and Neptune making the assumption of a thermally conducting boundary layer between the hydrogen-helium-dominated outer envelope and the metal-rich inner envelope. We investigated the influence of three parameters: layer thickness $\Delta m$, thermal conductivity $\lambda$, and the onset time on the planets' cooling behaviour. \\

We find that the inclusion of even a thin region that is permanently thermally conductive impacts the planet's cooling significantly, first strongly accelerating the surface cooling, and then slowing it down considerably. This leads several of the considered evolutionary tracks to cross the adiabatic comparison case. It opens up the possibility of a planet with such a TBL appearing brighter (Neptune) or fainter (Uranus) than an adiabatic one of  equal composition. The former case has similar results to those presented by \citet{Leconte13} for Saturn, who showed that a region of inhibited energy transport, in that case via layered convection, increases the present-day luminosity and thus prolongs the evolution. This proposed  mechanism can explain why adiabatic evolution models underestimate the luminosity of Saturn. \\

For higher TBL thicknesses, we find a plateau-like behaviour in the luminosity. This shows some resemblance to evolution calculations for  gaseous exoplanets, which appear overly large for their age, even if common heating contributions (Ohmic heating, tidal heating) are accounted for and thus need an additional heating term to match evolution models to their radius; WASP-39b \citep{Wakeford18,Poser19} is an example of this. This indicates that a TBL may also be relevant for some exoplanets.\\

For Uranus specifically, we find reproducing the observed present-day luminosity can only be achieved by an interface thick enough to trap the majority of its primordial heat inside the planet so that the outer envelope hits equilibrium with the irradiation of the sun. This can, and usually does, happen at a time previous to the present day, a scenario also explored in \citet{Nettelmann16}. Under these circumstances, the planet's interior remains as hot as in the beginning and a temperature step of $\sim \SI{8000}{K}$ develops between outer and inner envelope. The high central temperature of $> \SI{2e4}{K}$ is similar to that found in the Uranus model no. 2 by \citet{Vazan20}, which also features a steep metallicity increase in the envelope. This hot interior would also preclude the water in the inner envelope from reaching the superionic phase \citep{French16}, which would prohibit scenarios such as a substantial 'frozen'{}\ interior explored by \citet{Stixrude20}, for example. However, for certain configurations a stably stratified region still develops in Uranus' interior, which might account for its magnetic field morphology \citep{Stanley04}, for which a superionic region in the mantle has been put forward as a solution in the past \citep{Redmer11}. \\

Solutions for Neptune fall into two disjunct classes that occur across  the entire investigated range of $\Delta m$ and $\lambda$ values, although they do require certain combinations thereof. 
Neptune today could be in phase III --as our models predict for Uranus-- if the layer thickness exceeds \SI{15}{km}. In that case, the thermal state and interior of Uranus and Neptune could be similar: a hot (10,000-20,000 K) interior, which is either slowly cooling (Neptune) or slowly warming up (Uranus).\\
Neptune could also be in phase IV if $\Delta m$ is narrower ($<\SI{15}{km}$) and the TBL of enhanced conductivity is ($\lambda/\lambda_{\rm H2O}\sim 100)$, perhaps suggesting a layered double-diffusive state. Equivalent models for Uranus would predict $T_\text{eff}$ be too high by 3--6~K. In that case, additional processes may play a role in both planets, which were neglected in this work.  Overall, this work suggests that the similar effective temperatures of Uranus and Neptune can be explained by similar thermal histories. A dichotomy --a new model for either planet \citep{Helled20b}-- remains a possible but unnecessary option.\\ 

Of course, the model presented here is a vast simplification. While we assumed a simple three-layer structure and rather thin TBLs, planet formation models for gaseous planets predict large-scale compositional gradients extending from the centre outward \citep{Vazan15}. The case of ice giants is less clear as planet formation models struggle to  predict the ice-to-gas ratio inferred for present Uranus and Neptune if they formed at their current orbital distances \citep{Frelikh17}. Moreover, a primordial moderate compositional gradient can also erode by convective mixing in Uranus \citep{Vazan20}, certainly contributing to its luminosity evolution. Indeed, we find that the top of the TBL in some models could be unstable against convection. On the other hand, this work shows that a thin TBL of enhanced conductivity in Neptune could be Ledoux-stable to convection and explain its brightness. In addition, condensation of the volatiles and latent heat release with the  associated faintness effect \citep{Kurosaki17} certainly occurs in the atmosphere to some extent, which we did not include in our calculations. 
Moreover, the evolution models all start from a hot, adiabatic model of $T_\text{bar}=\SI{700}{K}$. However, the initial energy budget affects the evolution. Tying the evolution models to formation scenarios will be the subject of further work.

This work does not constrain the physical mechanism by which such a TBL would develop. If it formed later in the planet's lifetime, the influence on retaining heat in the interior is weaker. For example, Neptune's interior might only be several hundred~K warmer than the adiabatic case. Our models are consistent with the proposed possible water-hydrogen demixing if the TBL occurs underneath such a region. Indeed, because the TBL would accelerate the cooling of the outer envelope and create more favourable conditions to demixing than a prior adiabatic evolution would, a TBL that formed through a mechanism other than H$_2$-H$_2$O demixing could help facilitate the sedimentation. 
Ultimately, the existence of a TBL remains an open question; nevertheless, it is a powerful mechanism to explain the luminosity of Uranus and Neptune within a coherent framework.
\begin{acknowledgements}
We thank the referee for helpful comments for this publication. We thank Anna Julia Poser, Martin French, and Armin Bergermann for fruitful discussions. This project was funded by the German Research Foundation DFG as part of the Research Unit FOR-2440 \glqq Matter Under Planetary Interior Conditions \grqq. LS also thanks the International Space Science Institute in Bern for its hospitality.
\end{acknowledgements}
\bibliographystyle{aa}
\bibliography{Literature.bib}
\end{document}